\documentclass{JHEP3}

\usepackage{epsfig,multicol,bbm}
\newcommand{\bmat}{\left(\begin{array}}
\newcommand{\emat}{\end{array}\right)}
\newcommand{\be}{\begin{equation}}
\newcommand{\ee}{\end{equation}}
\newcommand{\bea}{\begin{eqnarray}}
\newcommand{\eea}{\end{eqnarray}}
\newcommand{\nn}{\nonumber}
\newcommand{\eq}[1]{Eq.~(\ref{#1})}

\def\lsim{\raise0.3ex\hbox{$\;<$\kern-0.75em\raise-1.1ex\hbox{$\sim\;$}}}
\def\gsim{\raise0.3ex\hbox{$\;>$\kern-0.75em\raise-1.1ex\hbox{$\sim\;$}}}
\def\vanish{\raise1ex\hbox{$\hspace*{-1cm}\nearrow^0$}}

\newcommand{\dmd}{\Delta m_{B_{d}}}  
\newcommand{\dms}{\Delta m_{B_{s}}}  

\def\Mp{M_{\rm{P}}}

\def\be{{\mathbf{e}}}

\def\nn{\nonumber}

\def\sn{\rm{s}}
\def\cn{\rm{c}}
\def\tn{\rm{t}}

\newcommand\fverb{\setbox\pippobox=\hbox\bgroup\verb}
\newcommand\fverbdo{\egroup\medskip\noindent%
                        \fbox{\unhbox\pippobox}\ }
\newcommand\fverbit{\egroup\item[\fbox{\unhbox\pippobox}]}
\newbox\pippobox

\title{$\mathbf{SO(10)}$ unified models and soft leptogenesis}
\author{E. J. Chun\\
Korea Institute for Advanced Study,\\
207-43 Cheongryangri-dong, Dongdaemun-gu,\\
Seoul 130-012, Korea \\
\email{ejchun@kias.re.kr}\\
{and}\\
Michigan Center for Theoretical Physics\\
Department of Physics, University of Michigan\\
Ann Arbor MI 48109, USA }
\author{L. Velasco-Sevilla\\
        W. I. F. Theoretical Physics Institute,\\
        University of Minnesota, \\
        116 Church St SE, Minneapolis, MN 55454, USA\\
        \email{liliana@physics.umn.edu} }

\preprint{FTPI-MINN 07/02\\ UMN-TH-2535/07 \\ MCTP-07-08}

\abstract{Motivated by the fact that, in some realistic models
combining $SO(10)$ GUTs and flavour symmetries, it is not possible
to achieve the required baryon asymmetry through the CP asymmetry
generated in the decay of right-handed neutrinos, we take a fresh
look on how deep this connection is in $SO(10)$. The common
characteristics of these models are that they use the see-saw with right-handed neutrinos,
predict a normal hierarchy of masses for the neutrinos observed in
oscillating experiments and in the basis where the right-handed
Majorana mass is diagonal, the charged lepton mixings are tiny.
 In addition these models link the up-quark Yukawa matrix to the neutrino Yukawa matrix $Y^\nu$ with the special feature of $Y^\nu_{11}\to 0$.
Using this condition,
we find that the required baryon asymmetry of the Universe can be explained by the
soft leptogenesis  using the  soft $B$ parameter of the second
lightest right-handed neutrino whose mass turns out to be around
$10^8$ GeV.  It is pointed out that a natural way to do so is to
use no-scale supergravity where the value of $B\sim1$ GeV is set through
gauge-loop corrections.}


\begin{document}

\section{Introduction}
 Some of the unanswered questions left by the Standard Model (SM) acquire
 a new light in Grand Unified Theories (GUTs).
Among them the understanding of the structure of masses has an
important role. Theories based on $SO(10)$ seem to have a special
place due to the possible structure of masses that can be
achieved. First of all the spinorial ${\mathbf{16}}$
representation can accommodate all the known fermions
including  a right handed neutrino component, $N$, which makes it
possible to embed in a natural way a see-saw mechanism for the
explanation of the tiny masses of low energy neutrinos. On the
other hand baryogenesis through leptogenesis \cite{FY} is a simple
mechanism to explain the observed baryon asymmetry in the
universe. Here a lepton asymmetry is dynamically generated and
then converted into a baryon asymmetry due to $B+L$ violating
sphaleron interactions.
The lightest of the right-handed neutrinos is produced by thermal
scattering after inflation. It subsequently decays
out-of-equilibrium to a lepton and a Higgs doublet producing a CP
and lepton number violating asymmetry.   The connection of
leptogenesis to $SO(10)$ GUTs is then natural, both by the inclusion
of right handed neutrinos in the ${\mathbf{16}}$s and by the possible
variation of their masses which can be some orders of magnitude below $M_{\rm{GUT}}\sim
10^{16}$ GeV.

 However in realistic models where the expansion parameter
 describing the Yukawa couplings of neutrinos,
$\epsilon_{\nu}$ is of the order of the  expansion parameter
describing the Yukawa couplings of up-type quarks, $\epsilon_u$
\cite{Ross:2002fb}, \cite{Chen:2004xy} \cite{Bando:2004hi}, the
value needed for the mass of the lightest right-handed neutrino,
$M_{N_1}\sim 10^{(7-8)}$ GeV, lies below the bounds for a successful
thermal leptogenesis;
 $M_{N_1}\gsim 10^{9}{\mathrm{GeV}}$ \cite{Davidson:2002qv}.
This prompts us to address two questions as follows:

\noindent a) How general is this statement in the context of Grand
Unified models where one naturally gets
\bea \epsilon_{\nu}\sim \epsilon_u \sim \sqrt{\frac{m_u}{m_c}}\; ?
\label{eq:epnu_epu} \eea
b) If that is the case for many classes of such models, what are
the possible  scenarios for leptogenesis that we may require to
consider in order to preserve such a feature of Grand Unified
Models?

In the present work we show how easy it is to generate the
relation \eq{eq:epnu_epu} and at the same time obtain $M_{N} \sim
10^{(7-8)}$ GeV for lighter right-handed neutrinos.  This
will be shown explicitly in the limiting case of the vanishing
neutrino mixing angle associated with reactor experiments
$s_{13}\rightarrow 0$, and of the dominant contribution of two
light right-handed neutrinos. The contribution from non-zero $s_{13}$ and the heaviest
right-handed neutrino can be taken  as perturbations.
This enables us to determine, from the current
experimental values, the possible form of the Dirac couplings of
left-handed neutrinos in the basis where the mass of the
right-handed neutrinos is diagonal, but without assuming any
particular hierarchy among them.

We then embed these results in a $SO(10)$ context,  without fully
specifying a model but rather making choices that are compatible
with GUTs, which can be used as a starting point in the construction
of a complete $SO(10)$ model.  In fact our results are compatible
with the models of \cite{Ross:2002fb} (RV), \cite{Chen:2004xy} (CM),
\cite{Bando:2004hi} (BKOT), and \cite{Dermisek:2006dc}  (DHR), in which
one considers symmetric matrices at the
scale where $SO(10)$ has not been broken and then explains the
generation of fermion masses with a minimal content of Higgs bosons, such as
two $\mathbf{10}$ representations, which are the Higgs bosons in the
$u$ and $d$ sectors, a $\overline{\bf 126}$ representation to
generate masses for right-handed neutrinos and a
non-renormalizable operator $\mathbf{45}$ to distinguish some
features of the fermion masses. These are generic features that
may be used to construct more specific models, and in fact have
been used extensively (\cite{Ross:2002fb}-\cite{Chen:2003zv}).

The neutrino Yukawa texture and the right-handed neutrino masses
determined in many works do not allow for the standard leptogenesis
\cite{FY} to produce the required baryon asymmetry of the
Universe. Other mechanisms of inducing the appropriate baryon
asymmetry in extensions of the MSSM may be implemented (e.g.
\cite{Campbell:1992hd} and see \cite{Mohapatra:2005wg} for a
review) or also mechanisms just using extra right-handed neutrinos (e.g. \cite{Senoguz:2007hu}).

 However, we observe that the soft leptogenesis of
\cite{D'Ambrosio:2003wy}, allowing a resonance condition with a
small $B$ term, works quite well within our scenario through the
decay of the second lightest right-handed sneutrino whose mass is
around  $10^8$ GeV.  We will see that the required small $B$ term
($B\sim 1$ GeV) arises from the gauge one-loop correction
involving a heavy GUT gaugino once its tree-level value vanishes
as it can be arranged easily in no-scale supergravity.

\section{Possible forms of ${m^\nu_{LR}}$ and $M_R$ in models with an underlying Grand Unified theory}
We will identify elements of the right-handed neutrino mass matrix
$M_R$ and the Dirac neutrino mass matrix, $m^\nu_{LR} = Y^\nu v
\sin\beta/\sqrt{2}$, which are related to the low energy
observables (neutrino masses and mixing angles) by
\bea
m^{\nu}=U^T \left[\begin{array}{ccc}
m_{\nu_1} &           &\\
          & m_{\nu_2} &\\
          &           & m_{\nu_3}
\end{array}
\right] U^{*}= -m^\nu_{LR}M^{-1}_{R}(m^\nu_{LR})^T .
\label{mass-relations}
 \eea
Here $U$ is the neutrino  mixing matrix and $m_{\nu_i}$ are the
neutrino mass eigenvalues. This expression is valid in the basis
where charged leptons are diagonal, if their matrix is not
diagonal then we get $U=U^{\nu}U^{e *}$.  Following the standard
parameterization, let us write
\bea
U=U_{23} P_\delta^*
U_{13}P_\delta U_{12} P_m
\eea
 where $U_{ij}$ is the rotation matrix in the ($i,j$) plane,
  $P_\delta = {\rm
diag}[e^{-i\delta},1,e^{i\delta}]$ and $P_m={\rm
diag}[e^{i\rho},e^{i\sigma},1]$ are the Dirac and Majorana phase
matrices, respectively.  We will use the notation
$(m^\nu_{LR})_{ij} =m_{ij}$ and $(Y^\nu)_{ij}=y_{ij}$, that is,
$m_{ij}=y_{ij} v \sin\beta/\sqrt{2}$.

For numerical values, we use the latest values of the fits for
neutrino oscillation observables \cite{Gonzalez-Garcia:2004jd}:
\bea
\Delta m^2_{21}&\in& [7.3,9.3]\times 10^{-5} {\mathrm{eV}^2},\quad t^2_{12}\in [0.28,0.60]\nn\\
\Delta m^2_{32}&\in& [1.6,3.6]\times 10^{-3} {\mathrm{eV}^2},\quad t^2_{23}\in [0.25,2.1]\nn\\
&& \hspace*{3.85cm} s^2_{13} \leq 0.041. \label{eq:oscill_inp}
\eea
Following the GUT relations consistent with the hierarchical
pattern of all the fermion masses, we will work with the normal
hierarchy;
\bea m_{\nu_3} \gg m_{\nu_2} \gg m_{\nu_1}.
\label{eq:assumhierch} \eea
 Then, we make the following definitions to help us trace the
hierarchy of low-energy neutrino observables:
\bea {\bf r}\equiv \sqrt{\frac{\Delta m^2_{21}}{\Delta
m^2_{32}}}={m_{\nu_2}\over m_{\nu_3}},\quad {\bf t} \equiv
\frac{m_{\nu_1}}{m_{\nu_3}}.\label{eq:def_r_t} \eea

\subsection{GUT considerations\label{sec:gutconsid}}
Let us recall that the Yukawa sector of $SO(10)$ at the
renormalizable level comes from the following allowed couplings in
the matter Lagrangian \cite{Mohapatra:1986bk};
\bea {\mathcal{L}}_{M}= y_{ij}^{10} ({\bf 16})_{i} ({\bf 16})_{j}
({\bf 10}) + y_{ij}^{120} ({\bf 16})_{i} ({\bf 16})_{j} ({\bf
120}) + y_{ij}^{126} ({\bf 16})_{i} ({\bf 16})_{j} (\overline{\bf
126}) \eea due to the famous decomposition ${\bf 16} \otimes {\bf
16} = {\bf 10}_s\oplus {\bf 120}_s \oplus \overline{\bf 126}_s$.
As has been widely stated, the minimal Higgs content in order to
generate Dirac masses for quarks and leptons and Majorana masses
for right-handed neutrinos is to have two Higgs bosons in the
$\mathbf{10}$ representation of $SO(10)$ and a Higgs in the
$\overline{\mathbf{126}}$ representation.

 The Yukawa couplings $y^{10}_{ij}$ associated with
 $\mathbf{10}$ representations give
\bea
(m^d)^T=m^e, \quad m^u=m^\nu_{LR},
\label{eq:mqreltions10}
\eea
while the  $\overline{\mathbf{126}}$  gives mass only to right
handed neutrinos.

Prior to the Super-Kamiokande experiments \cite{Fukuda:1998fd},
there were successful models which could reproduce fermion masses
with this Higgs content, however without explanation of the exact
relations between the elements of the same Yukawa matrix. With the
current data of the neutrino oscillation experiments
\cite{Gonzalez-Garcia:2004jd}, now it is clear that the minimal
content must be extended to fit neutrino masses. This can be done by
adding more $\mathbf{10}$, $\mathbf{120}$ or  $\mathbf{126}$ Higgs
fields or by adding non-renormalizable operators or other elements
in the theory beyond a GUT.

 If we consider non-renormalizable operators we can alter the structure of the Yukawa matrices.
The Lagrangian of these operators is
\bea {\mathcal{L}}_{M(ht)}={\mathcal {O}}^{ij} y^{{\mathcal
{O}}}_{ij},\quad \mathcal{O}_{ij} \equiv {\bf 16}_i\, {R_1\over
M_1}. . .  {R_k\over M_k}\, {\bf 10}\, {R_{k+1}\over M_{k+1}} . .
. {R_\ell\over M_\ell}\, {\bf  16}_{j}, \eea
where $R$'s are possible representations coupling to ${\mathbf{10}}$ and ${\mathbf{16}}$
and $y^{{\mathcal {O}}}_{ij}$ is their corresponding Yukawa
matrix.  An interesting case is the adjoint representation
${\mathbf{45}}$ whose vacuum expectation value (vev) can point to
any direction in the space spanned by the ${\mathbf{45}}$ generators as long
as it leaves the SM group $SU(3)_c \times SU(2)_L \times U(1)_Y$
unbroken. The space of ${\mathbf{45}}$ vev, that leaves unbroken the SM
group, lies in the two dimensional subspace of $U(1)'$ generators
of $SO(10)$ that commute with $SU(3)_c \times SU(2)_L \times
U(1)_Y$. For example when $SO(10)$ is broken to the $SU(3)_c\times
SU(2)_R\times SU(2)_L$ subgroup, the general form to which the
${\mathbf{45}}$ should be pointing is
\bea
\langle \mathbf{45} \rangle= (B-L +k T_{R 3})M_{\mathbf{45}}.
\eea
Successful and predictive scenarios
can be obtained when incorporating flavour symmetries into GUTs. With a clever choice
of a flavour symmetry, it may not be
needed to invoke additional Higgs fields beyond two $\mathbf{10}$s
and one $\overline{\mathbf{126}}$ representations. However
majority of such examples do require at least one
non-renormalizable operator (see \cite{Albright:2002np} for a
review).

From the experimental information on $V_{\rm{CKM}}$ and the quark
masses we can reconstruct the possible forms that the quark mass
matrices, $m^q$, can acquire from the above considerations. However since we are just
able to construct the left diagonalizations of $m^q$ from
$V_{\rm{CKM}}$ we need to make more assumptions on $m^q$. One of
the most successful and well established assumptions is to
consider that the mixings in each quark sector are small such that
they remain small when combined in the $V_{\rm{CKM}}$. This
assumption can be naturally identified with a strong hierarchical
structure of $m^q$.

The diagonalizing matrix of $Y^q=\sqrt{2}m^q/v_q$ can be
parameterized as the multiplication of unitary matrices
$U_{23}U_{13}U_{12}$ where each $U_{ij}$ is  a rotation matrix in
the $(i,j)$ sector including some phases. When $Y^q$ is hierarchical,
the angles corresponding to this parameterization can be in fact
approximately identified with the  angles obtained from the approximate multiplication \cite{Roberts:2001zy},
$
U^\dagger_{12R}U^\dagger_{13R} U^\dagger_{23 R} Y U_{23 L} U_{13 L} U_{12 L}\approx Y_{\mathrm{diag}}$.
For the case of the up quarks, we have
\bea
y_u&\approx& |c^L_{12} c^R_{12} y^{\prime\prime}_{11} -e^{i\phi_L} c^L_{12}s^R_{12} y^{\prime\prime}_{12} - e^{i\phi_R} c^R_{12}s^L_{12} y^{\prime\prime}_{21}+s^R_{12} s^L_{12}y^{\prime\prime}_{22}|\nn\\
y_c&\approx& c^L_{12} c^R_{12} y^{\prime\prime}_{22} +e^{i\phi_L} c^L_{12}s^R_{12} y^{\prime\prime}_{12} +e^{i\phi_R} |c^R_{12}s^L_{12} y^{\prime\prime}_{21}+s^R_{12} s^L_{12}y^{\prime\prime}_{11}|,\nn\\
y_t&\approx & | y_{33}c^L_{23}c^R_{23}+c^R_{23}s^L_{23} y_{32} e^{i\phi_{23L}} + c^L_{23}s^R_{23} y_{23} e^{i\phi_{23R}}  +s^L_{23}s^R_{23}y_{22}|
\eea
where $y^{\prime \prime}$ is the matrix after the unitary transformation in the $(2,3)$ and $(1,3)$ sectors.
We are particularly interested in analyzing the behavior in the $(1,2)$ sectors.
For small rotations we have $c^R_{12}=c^R_{21}\approx 1$,
 $s^R_{12}=y^{\prime \prime}_{21}/y^{\prime \prime}_{22}$
  and $s^L_{12}=y^{\prime \prime}_{12}/y^{\prime \prime}_{22}$.
 When $y^{\prime\prime}_{11} \ll y^{\prime\prime}_{12} y^{\prime\prime}_{21}/ y^{\prime\prime}_{22}$, one gets
\bea
|y_u|\approx   |s^L_{12} s^R_{12} y^{\prime \prime}_{22} |\quad
|y_c|\approx   | y^{\prime \prime}_{22}|,
\label{eq:yuandyc}
\eea
which, for the symmetric case, gives the famous relation $s^u_{12}\approx \sqrt{{m_u}/{m_c}}$.
This forms the basis of the Gatto-Sartori-Tonin   relation \cite{Gatto:1968ss}
\bea
V_{us}=\left|s^d_{12}-e^{i\Phi_1} s^u_{12}\right|\approx \left|\sqrt{\frac{m_d}{m_s}}-e^{i\Phi_1} \sqrt{\frac{m_u}{m_c}} \right|.
\label{eq:gst}
\eea
This relation is in good agreement (originally the
contribution just from the down sector  was assumed, but now the
contribution from the up sector has become important) with the
experimental value of $V_{us}$. In order to review the conditions
for which $y^{\prime\prime}_{11}$   $\ll y^{\prime\prime}_{12}
y^{\prime\prime}_{21}/ y^{\prime\prime}_{22}$, let us give
explicitly their expressions in  terms of the small rotation
angles and the original Yukawa couplings:
\bea
y^{\prime\prime}_{11} &\approx&y_{11}-\frac{[s^R_{23}y_{12} e^{i\phi^R_{23}}+c^R_{23} y_{13} ]
[ s^L_{23} y_{21} e^{i\phi^L_{23}}+ c^L_{23} y_{31}] }{y_{33}c^L_{23}c^R_{23}+ O(y_{23}y_{32}/y_{33})}\nn\\
\frac{y^{\prime\prime}_{12} y^{\prime\prime}_{21}}{ y^{\prime\prime}_{22}} &\approx &
\frac{[c^R_{23} y_{12}-s^R_{23} y_{13}e^{-i\phi^R_{23}}][c^L_{23}y_{21}-s^L_{23}y_{31}e^{-i\phi^L_{23}}  ]}
{y_{22}c^L_{23}c^R_{23}-c^R_{23} s^L_{23} y_{32} e^{-i\phi^L_{23}}- c^L_{23}s^R_{23} y_{23}e^{-i\phi^R_{23}}+s^L_{23}s^R_{23}y_{33}}.
\label{eq:explicity1j}
\eea
From these expressions it is clear that, if the original Yukawa
matrix contains a zero in the $(1,1)$ position, then the inequality
$y^{\prime\prime}_{11} \ll y^{\prime\prime}_{12}
y^{\prime\prime}_{21}/ y^{\prime\prime}_{22}$ is immediately
achieved because of $y_{33}\gg y_{22}$ leading to the desired
hierarchy between charm and top quark masses. However notice that
$y_{11}$  does not have to be exactly zero.  As long as it is
suppressed enough with respect to $y^{\prime\prime}_{12}
y^{\prime\prime}_{21}/ y^{\prime\prime}_{22}$ we can always have a
relation of the type \eq{eq:gst}.  This condition is referred hereafter
as the limit $m^u_{11}\rightarrow 0$. As we have seen it can be realized in a particular basis
of the Yukawa couplings satisfying \eq{eq:yuandyc},  which is however
a basis independent statement.

For a symmetric case, if we assume $y_{12}\ll y_{23}$ then in any
of the cases of having $y_{22}\sim y_{23}$, $y_{22}\ll y_{23}$ or
$y_{22} > y_{23}$, the second term in the first of the equations
of \eq{eq:explicity1j} will be smaller than the expression for
$y^{\prime\prime}_{22}$. Thus the interesting case is when
$y_{11}$ is the leading contribution in $y^{\prime\prime}_{11}$.
Let us take the case $y_{22}\sim y_{23}$ then the leading
contribution in $y^{\prime\prime 2}_{12}/y^{\prime \prime}_{22}$
will be simply $y^2_{12}/y_{22}$ and  hence we obey the condition
$y^{\prime\prime}_{11} \ll y^{\prime\prime}_{12}
y^{\prime\prime}_{21}/ y^{\prime\prime}_{22}$, in terms of the
original matrix elements: $y_{11}\ll y^2_{12}/y_{22}$.

For matrices of the form \cite{Roberts:2001zy},
\cite{Ross:2002fb}, \cite{Ross:2004qn}, where the elements
$Y^u_{12}$ and $Y^u_{22}$ are respectively $\epsilon^3_u$ and
$\epsilon^2_u$, one simply needs to require $Y^u_{11}\ll
\epsilon^4_u$. For matrices of the form \cite{Masina:2006ad},
which are also symmetric and hence compatible with $SO(10)$, one
recovers the requirements of the element $Y^u_{11}$ with the
present analysis. In this texture, $Y^u_{12}$ is $\epsilon^3_u
\sim \lambda^6 $, $Y^u_{22}$ is zero but $Y^u_{23}$ is
$\epsilon^2_u$, and hence $(y'')^2_{12}/y''_{22}=\epsilon^4$.
Thus, by making  $Y^u_{11}=O(\epsilon^4_u)$ in this case, an
$O(1)$ correction to the relation $s^u_{12}\approx \sqrt{m_u/m_c}$
can be made.

\subsection{Compatibility with the experimental information}

Symmetric fits to the quark masses can be used as a guideline to
construct models with underlying $SO(10)$, or $SU(4)_c\times
SU(2)_L\times SU(2)_R$, which after the $SO(10)$ breaking assume a
symmetric structure. However a small departure from symmetric
matrices do not change the qualitative behavior of such fits and
can be made compatible with $SO(10)$ and flavour symmetries.

The fits of these matrices into the experimental information can
be made in many ways depending on our theoretical assumptions. A
minimal choice is to assume that the supersymmetric corrections to
the quark masses will not have a strong impact on the {\it ratio}
of masses that we use for the fit. We also assume that the Yukawa
matrices are the only source of CP violation and that the
contributions from the transformation of the squared soft mass
matrices of the supersymmetric particles to the basis where the
Yukawa matrices are real and diagonal are negligible. We specify
this last requirement because we can choose to fit the quark mass
matrices with ratios of masses and with the fits of the unitary
triangle, where various CP violating experiments are taken into
account.

With increasing precision in the determination of the fits of the
unitary triangle in the SM, however, one has a very tight
constraint on the parameters and one should not regard this as a
final fit of a particular texture but as an indication of the
current compatibility of such texture with the theoretical
assumptions and experimental information. The purpose of the
present analysis is to clarify the consequences of having a
negligible $Y^u_{11}$ entry, giving the relation $s^u_{12}\approx
\sqrt{m_u/m_c}$, and furthermore a low range of right handed
neutrino masses, which turn out to be incompatible with the
standard thermal leptogenesis to produce the observed baryon
asymmetry of the universe. For completeness we present in Appendix (\ref{app:fit}) the current fit of a symmetric texture with negligible
$Y^u_{11}$ element, compatible at 68\% C.L. with up-to-date fits of the unitary
triangle.

It is interesting to see that two different choices of symmetric
matrices in the up sector, (\cite{Ross:2002fb},
\cite{Chen:2004xy}, \cite{Bando:2004hi}, \cite{Dermisek:2006dc})
and (\cite{Masina:2006ad}, \cite{Masina:2006pe}), give rise to
different phenomenology and consequences, e.g., for leptogenesis.
This is because we then have clearer selection criteria on how to single out the
models, apart from direct fittings to experimental mass matrices
and quark mixing. Further precision analyses in the unitary
triangle will constrain more such possibilities.

For non-symmetric fits \cite{Kane:2005va}, we have a rather
different situation because often $m^q_{11}$ should be of the
order of $m^q_{12}$ to reproduce appropriate mixings and quark
masses. The non symmetric fits can be used to construct models
with underlying $SU(5)$ symmetries (see for example
\cite{Altarelli:2000fu}).

In the following, we will consider the limit
$(m^\nu_{LR})_{11}=m^u_{11}\rightarrow 0$ and  explore its
consequences for neutrino oscillations and leptogenesis.
\subsection{Consequences of $(m^\nu_{LR})_{11}\rightarrow 0$ \label{sec:casemdeq0}}
\subsubsection{General structure of the matrix of low energy neutrinos}
The explicit form of  the low energy neutrino mass components from
Eq.~(\ref{mass-relations}) is
\bea
\frac{m^\nu_{ee}}{m_{\nu_3}}&=&
{\mathbf{r}} c^2_{13} s^2_{12}+ e^{2i(\delta-\sigma)}s^2_{13}+ {\mathbf{t}}c^2_{13} c^2_{12}e^{-2i\rho} \nn\\
\frac{m^\nu_{e\mu}}{m_{\nu_3}}&=&
s_{23}s_{13}c_{13}e^{i(\delta-2\sigma)} + {\mathbf{r}} c_{13}s_{12}(c_{12}c_{23}- \frac{s_{12}s_{13}s_{23}}{e^{i\delta}})
- \frac{{\mathbf{t}}}{e^{2i\rho}}c_{12}c_{13}(c_{23}s_{12}+\frac{c_{12}s_{13}s_{23}}{e^{i\delta}})\nn \\
\frac{m^\nu_{e\tau}}{m_{\nu_3}}&=&
c_{23}c_{13}s_{13}e^{i(\delta-2\sigma)} - {\mathbf{r}}c_{13}s_{12}(\frac{c_{23}s_{12}s_{13}}{e^{i\delta}}+c_{12}s_{23} )
+\frac{{\mathbf{t}}}{e^{2i\rho}}c_{12}c_{13}(c_{23}s_{12}-\frac{c_{12}s_{23}s_{13}}{e^{i\delta}})\nn\\
\frac{m^\nu_{\mu\mu}}{m_{\nu_3}}&=&
s_{23}^2c_{13}^2e^{-2i\sigma}+{\mathbf{r}}(c_{12}c_{23}-\frac{s_{12}s_{13}s_{23}}{e^{i\delta}})^2
-\frac{\mathbf{t}}{e^{2i\rho}}(c_{23}s_{12}+c_{12}s_{13}s_{23})^2\nn\\
\frac{m^\nu_{\mu\tau}}{m_{\nu_3}}&=&
\frac{c_{23}s_{23}c_{13}^2}{e^{2i\sigma}}-{\mathbf{r}}(\frac{c_{23}s_{12}s_{13}}{e^{i\delta}}+c_{12}s_{23})(c_{12}c_{23}-\frac{s_{12}s_{13}s_{23}}{e^{i\delta}})\nn\\
&-&\frac{\mathbf{t}}{e^{2i\rho}}(c_{23}s_{12}+\frac{c_{12}s_{23}s_{13}}{e^{i\delta}})(s_{12}s_{23}-\frac{c_{12}c_{23}s_{13}}{e^{i\delta}})\nn\\
\frac{m^\nu_{\tau\tau}}{m_{\nu_3}}&=&
\frac{c_{13}^2c_{23}^2}{e^{2i\sigma}}+{\mathbf{r}}(\frac{c_{23}s_{12}s_{13}}{e^{i\delta}}+c_{12}s_{23})^2 + \frac{\mathbf{t}}{e^{2i\rho}}(s_{12}s_{23}-\frac{c_{12}c_{23}s_{13}}{e^{i\delta}})^2,
\label{eq:completeform}
\eea
where we have expressed the elements of $m^\nu$ in terms of the
sub-indices $e$, $\mu$ and $\tau$. In Appendix
(\ref{sec:appn_um_mnu}) we have written the numerical central
values of the angles, up to $\mathbf{t}$, $s^2_{13}$ and possible phase variations.
From
\eq{eq:numericmnu} we can see that in the limit of $s_{13}=0$, all
the numerical entries that are not multiplied by $\mathbf{t}$ in
\eq{eq:completeform} do not change its order of magnitude. On the
other hand the complete form of $m^{\nu}$ in terms of a diagonal
matrix $M_R=\mbox{diag}[M_1,M_2,M_3]$ and a general matrix
$m^\nu_{LR}$ is given by
\bea m^{\nu} =& & \sum_i \frac{1}{M_i} \left[
\begin{array}{ccc}
m^{  2}_{1i}&  m^{ }_{1i} m^{ }_{2i}  &  m^{ }_{1i}m^{ }_{3i}\\
m^{  }_{1i} m^{ }_{2i}   &  m^{  2}_{2i}  &  m^{ }_{2i}m^{ }_{3i}\\
m^{  }_{1i} m^{ }_{3i}  &  m^{ }_{2i} m^{ }_{3i}  &  m^{  2}_{3i}
\end{array}
\right] .
\label{eq:mn_MR_md}
\eea

Now when $(m^\nu_{LR})_{11}=m_{11}\rightarrow 0$, this matrix
acquires a very simple form. Then we can simply identify the
elements of \eq{eq:completeform} with \eq{eq:mn_MR_md} and find
the form of $M_i$ in terms of $m_{ij}$ and the restrictions of its
elements.

It is clear from \eq{eq:completeform} that the contribution from
${\bf t}$ can become relevant just for
 $m^{\nu}_{ee}$ and $m^{\nu}_{e\tau}$. But since ${\mathbf{t}}< O(0.1)$ according to \eq{eq:oscill_inp},
 this contribution can be at most as the same order of the rest of the contributions in $(m_{\nu})_{11}$ or $(m_{\nu})_{13}$.
  From \eq{eq:mn_MR_md} we can see that these elements are given by
\bea
m^{\nu}_{ee}=\frac{m_{12}^2}{M_2}+\frac{m^2_{13}}{M_3},\quad
m^{\nu}_{e\tau}=\frac{m_{12}m_{32}}{M_2}+\frac{m_{13}m_{33}}{M_3}.
\eea
Identifying certain elements in $m^\nu_{LR}$, we can obtain the predicted
ranges for the right-handed neutrino masses which will be derived in detail
in the next subsection.  In this subsection, we try to extract the
general expressions for the parameters such as
 \bea
 M_1,\quad \frac{M_1}{M_2}\quad {\mathrm{or}}\quad
\frac{M_1}{M_3} ,\quad {\mathrm{and}}\quad
\tilde{m_1}=\frac{(Y^{\nu\dagger}Y^\nu )_{11} v^2}{M_{1}},
\label{eq:lepto_param}
 \eea
which are relevant for leptogenesis. Note that
$m^{\nu}_{ee}$ is the element which contains less parameters and
so we can make less assumptions when deriving expressions for \eq{eq:lepto_param}.
Then we have
 \bea
M_2&=&\frac{m^2_{12}[1+\frac{M_2}{M_3}\frac{m^2_{13}}{m^2_{12}}]}{m_{\nu_3}[
{\mathbf{r}} c^2_{13} s^2_{12}+ e^{2i(\delta-\sigma)}s^2_{13}+
{\mathbf{t}}c^2_{13} c^2_{12}e^{-2i\rho}]}\nn \\
& \gtrsim&  2\times 10^{16}
y^2_{12}\frac{\sin^2\beta}{2}\left[1+\frac{M_2}{M_3}
\frac{m^2_{13}}{m^2_{12}}\right]\ {\mathrm{GeV}}, \label{eq:M2complt}
 \eea
where we have set the bound on $M_2$ by taking the numerical
values of the first two contributions of the denominator in
\eq{eq:M2complt}.

From this
relation we can study the behavior of the parameters in terms of
\bea
a\equiv \frac{M_2}{M_3}\frac{m^2_{13}}{m^2_{12}}.
\eea
For $a\ll 1$ we can see that the order of $M_2$ is determined just
by $y^2_{12}$. In this case, considering the expressions of
$m^\nu_{\mu\mu}+m^\nu_{\tau\tau}$, we obtain the ratio of $M_1$ to
$M_2$ given by
\bea
\frac{M_1}{M_2}=\frac{m^\nu_{ee}(m^2_{21}+m^2_{31})}{(m^\nu_{\mu\mu}+m^\nu_{\tau \tau})m^2_{12}(1+a)-m^\nu_{ee}
\left[a(m^2_{23}+m^2_{33})\frac{m^2_{12}}{m^2_{13}}+(m^2_{32}+m^2_{22}) \right] } .
\eea
Then, we find that generic $SO(10)$ models lead to $\tilde{m}_1$
given by
\bea
%
&&\tilde{m}_1 = \mbox{max}[m^\nu_{\mu\mu}+m^\nu_{\tau\tau}, b\
m^\nu_{ee}]\,,
\quad \mbox{ where}\nonumber\\
&& b\equiv
\frac{\left[a(m^2_{23}+m^2_{33})\frac{m^2_{12}}{m^2_{13}}+(m^2_{32}+m^2_{22})\right]
}{m^2_{12}(1+a)} \;.  \label{eq:tminM2}
 \eea
We can consider the cases for
$ b\lesssim 1\; {\mathrm{and}}\; b\gg1$.
For $b\lesssim 1$, the order of magnitude of $\tilde{m}_1$ is fixed simply by
$m_{\mu\mu}+m_{\tau\tau} \sim (10^{-2},10^{-1})$ eV.  For $b\gg 1$, $\tilde{m}_1\sim
b(10^{-3},10^{-2})$ eV. 
 This points out an important consequence that
the right-handed neutrino gets out of thermal equilibrium when it
is very non-relativistic [see next section] because
$m_* < \tilde{m}_1 $ where
\bea
 m_*=\frac{16 \pi^{5/2}}{3\sqrt{5}} g^{1/2}_*
\frac{v^2}{\Mp}\simeq 1.6\times10^{-3} \mbox{ eV}
 \eea
 with $g_*=225$ for the relativistic degrees of freedom in
supersymmetric standard model.
This brings  a strong suppression  to the CP asymmetry generated
through the decay of the right-handed neutrino.

 For $a\gg 1$, instead of the bound for $M_2$, we
determine the bound for $M_3$, which can be obtained by taking the
replacements: $M_2 \leftrightarrow M_3$ and $m_{13}\leftrightarrow
m_{12}$ in Eq.~(\ref{eq:M2complt}).
 Similarly, the corresponding $\tilde{m}_1$ is given by the
 relation (\ref{eq:tminM2}) with exchanging the indices $2\leftrightarrow3$
 for the expression of $b$.

This has the same behavior as \eq{eq:tminM2} except that in GUT
models the natural assumption is to
 have $y^\nu_{33}=O(1)$ and hence the second factor $b m^\nu_{ee}$ is likely to be the
 dominant.  Then this contribution to $\tilde m_1$ goes like $\tilde m_1 \sim m_{ee} y^{\nu
2}_{33}/y^{\nu 2}_{13}$
  which could be significantly bigger than $m_*\sim 10^{-3}$ eV.

In all the models \cite{Ross:2002fb}, \cite{Chen:2004xy}, \cite{Bando:2004hi} and \cite{Dermisek:2006dc}
 in the basis where $M_R$ is diagonal{\footnote{For the models \cite{Bando:2004hi}, \cite{Chen:2004xy},
  this transformation has been performed since for those models  $M_R$ is not diagonal in the basis
  where the underlying symmetry is broken.}} the condition \eq{eq:yuandyc} is satisfied
  for the case of the Dirac coupling of neutrinos and  then
  we can identify its behavior in terms of $a$ and $b$.
In \cite{Chen:2004xy}, a particular realization of the case with
$a\gg 1$ and  $y^\nu_{3,3}=1$ was explored and one gets $M_1\sim
10^7$ GeV and the wash-out factor, $\sim 10^{-5}$, to the leptonic
CP asymmetry in the decay of right-handed neutrinos. The other
parameters can bee seen in Table \ref{tbl:so10mods}.
 In \cite{Ross:2002fb}, one gets $a=0$ because effectively only two right-handed
neutrinos are taken into account and the mass of the lightest
right handed neutrino is $M_1=O(10^7)$ GeV. In \cite{Bando:2004hi},
$a\ll 1$ and $b\in (O(0.1),1)$ and the mass of the lightest
neutrino is $M_1=O(10^{(7,8)})$ GeV. A realization of the case
with $a\ll 1$ and $b \sim 1$ was explored in
\cite{Dermisek:2005ij} and \cite{Dermisek:2006dc} where
it is also not possible to achieve a successful thermal leptogenesis, due
to the washout factor, although the mass of the lightest right
handed neutrino is of order $M_1=O(10^{10})$ GeV.
In Table \ref{tbl:so10mods} we have summarized the properties
relevant for leptogenesis of these models and we will make more
comments about them in Section \ref{sec:so10realiz}.

In the next section we take the limit
\bea \sn_{13}\rightarrow 0,\quad \mathbf{t}\rightarrow 0 \quad and
\quad 1/M_3\rightarrow 0, \eea which gives $a\ll 1$ and $b\ll 1$
and allows to understand more deeply the  connection  of
$(m^\nu_{LR})_{11}\rightarrow 0$ with the low mass of $M_{1}$ in
GUT models.
%
%
%
%
\begin{table}[!ht]
\begin{center}
\begin{tabular}{|l l c l l l|}
\hline
Model    & Ref.                 & $M_R[{\mathrm{GeV}}]$       &  $a\equiv \frac{M_2}{M_3}\frac{m^2_{13}}{m^2_{12}}$    &  $b$  &$\tilde{m}_1$\\
\hline
BKOT    & $\!\!\!\!\!$ \cite{Bando:2004hi}  &{\small{$\!\!\left[\begin{array}{ccc}
                                              10^{(7,8)} &        &           \\
                                                           & 10^9   &           \\
                                                           &        & 10^{(14,15)}
                                               \end{array} \right]$}}     &$\begin{array}{c}
                                                                    {\mathrm{II.}}\  \sim 10^{-3}\\
                                                                    {\mathrm{III.}}\ \sim\times 10^{-4}\\
                                                                    {\mathrm{IV.}}\ \sim\times 10^{-4} \end{array}$  & $3\times 10^2 $     & $\begin{array}{l} b\ m^\nu_{ee} \sim\\ (10^{-1},1)\ {\mathrm{eV}} \end{array}$ \\
CM      & $\!\!\!\!\!$ \cite{Chen:2004xy}   & {\small{$\!\!\left[\begin{array}{ccc}
                                              10^7 &               &       \\
                                                   &  10^9  &      \\
                                                   &               & 10^{12}\\
                                               \end{array} \right]$}}  & $10$ & $10^{3}$ &  $\begin{array}{l} b\ m^\nu_{ee} \sim \\ (1,10)\ {\mathrm{eV}}\end{array}$ \\
DHR      & $\!\!\!\!\!$ \cite{Dermisek:2006dc} & {\small{$\!\!\left[\begin{array}{ccc}
                                              10^{10} &               &       \\
                                                   &  10^{12}  &      \\
                                                   &               & 10^{13}\\
                                               \end{array} \right]$}} & $10^{-5}$ & $O(1)$  & $\begin{array}{c}m^\nu_{\mu\mu}+m^\nu_{\tau \tau}\sim\\  m_{\nu_3}\lsim 0.625\ {\mathrm{eV}}\end{array}$ \\
RV       & $\!\!\!\!\!$ \cite{Ross:2002fb}     & {\small{$\!\!\left[\begin{array}{ccc}
                                              10^{7} &               &       \\
                                                   &  10^{8}  &      \\
                                                   &               & >10^{10}\\
                                               \end{array} \right]$}} & 0 & $O(1)$ & $m^\nu_{\mu\mu}+m^\nu_{\tau \tau}$  \\
\hline
\end{tabular}
\end{center}
\caption{\small{Models based on $SO(10)$ which do not generate the observed amount of baryon asymmetry through the decay of the right handed neutrinos in thermal leptogenesis.}}
\label{tbl:so10mods}
\end{table}
%
%
%
\subsubsection{Limit of $\mathbf{\sn_{13}\rightarrow 0}$, $\mathbf{t\rightarrow 0}$ and contributions proportional
to $\mathbf{1/M_3}$ negligible \label{sec:limits130}}
%
%
 The goal  of this analysis is to identify the possible shapes of $m^\nu_{LR}$ and $M_R$ in the basis where $M_R$
 is diagonal and the form of the mixing angles in terms of their entries. We will assume  that the
 mixing of charged leptons is small and hence can be ignored.  In
 the limit under consideration, Eq.~(\ref{eq:completeform})
 becomes much simpler as follows:
\bea
m_{\nu}=m_{\nu_3}\left[
\begin{array}{ccc}
s^2_{12} {\bf r}            & c_{12}c_{23}s_{12} {\bf r} & -c_{12} s_{12}s_{23}{\bf r}\\
c_{12}c_{23}s_{12}{\bf r}  & c_{12}^2c_{23}^2{\bf r}+s_{23}^2
e^{-2i\sigma} &
-c_{12}^2 c_{23}s_{23}{\bf r}+c_{23}s_{23}e^{-2i\sigma}\\
-c_{12}s_{12}s_{23}{\bf r} &  -c_{12}^2 c_{23}s_{23}{\bf r}+c_{23}s_{23}e^{-2i\sigma}  &c_{12}^2s_{23}^2{\bf r} + e^{-2i\sigma}c_{23}^2\\
\end{array}
\right] .
\label{eq:mn_low_info}
 \eea
When comparing the elements  $m^\nu_{ij}$ of \eq{eq:mn_low_info}
with those of \eq{eq:mn_MR_md}, we have six equations to solve,
which are more than the low-energy parameters to determine: one
mass ratio, two angles and one phase. Thus the elements of $m^D$ and $M_R$
are more restricted. By comparing the $m^\nu_{ee}$ component in
\eq{eq:mn_low_info} and \eq{eq:mn_MR_md}, we can readily identify
$M_2$:
\bea M_2= \frac{m_{12}^2}{m_{\nu_3} {\bf r} s_{12}^2} \,.
\label{eq:M2direct} \eea
Considering the ratio $m^\nu_{e\tau}/m^\nu_{e\mu}$,  we obtain an
important relation:
 \bea
 {t_{23}}=-\frac{m_{32}}{m_{22}} .
\label{eq:t23_13sec}
 \eea
Analogously the ratio $m^\nu_{\mu\tau}/m^\nu_{\tau\tau}$ leads to
 \bea
 {t_{23}}=\left[\frac{m_{21}m_{31}+\frac{M_1}{M_2}
m_{22}m_{32}}{m_{31}^2+ \frac{M_1}{M_2} m^2_{32}
}\right]\left[\frac{1+c^2_{12}t^2_{23}{\bf r} {\rm{e}}^{2i\sigma}
}{1-c^2_{12}{\bf r}{\rm{e}}^{2i\sigma}}\right].
\label{eq:t23_23sec}
 \eea
When we determine the ratio $M_1/M_2$ we can put a restriction on
the elements $m_{21}$ and $m_{31}$ from \eq{eq:t23_13sec} and
\eq{eq:t23_23sec}. Now the solar mixing angle is
given by the following simple equation;
\bea
{t_{12}}&=&\ \pm
\frac{m_{12}}{m_{22}}\frac{m_{32}}{\sqrt{m_{22}^2+m_{32}^2}}
\label{eq:tn_12_23} \eea
which is obtained by considering
$m^\nu_{ee}/m^\nu_{e\mu}$ and the relation $t_{23}^2=m_{32}^2/m_{22}^2$.
Adding $m^\nu_{\mu\mu}$ and
$m^\nu_{\tau\tau}$  we obtain
 \bea
 \frac{M_1}{M_2}=
\left[\frac{{\bf r }
s^{2}_{12}}{m^2_{12}}\right]\left[\frac{m^2_{21}+m^2_{31}}{{\rm{e}}^{-2i\sigma}
+{\bf r}
c^2_{12}(1-t_{23}^2 
 )}
\right] \approx
\left[\frac{{\bf r }
s^{2}_{12}}{m^2_{12}}\right](m^2_{21}+m^2_{31})e^{2i\sigma},
\label{eq:M1M2_23}
 \eea
 where the last equality follows from
\eq{eq:tn_12_23}. Analogously by adding  $m^\nu_{\mu\tau}$ and
$m^\nu_{\tau\tau}$ we obtain
 \bea
\frac{M_1}{M_2}=\left[ \frac{
 \left[\frac{{\bf r}s^{2}_{12}}{m^2_{12}}\right]
[m_{21}m_{31}+m^2_{31}]}{{\rm{e}}^{-2i\sigma}c_{23}(c_{23}+s_{23})
+{\bf r}c^{2}_{12}\left(
s_{23}(s_{23}-c_{23})-\frac{{t}^2_{12}}{m^2_{12}}[m^2_{32}+m_{22}m_{32}]
 \right)  }
\right] . \label{eq:M1M2_33}
 \eea
Now dividing \eq{eq:M1M2_23} by \eq{eq:M1M2_33} we can find
solutions for $p\equiv m_{21}/m_{31}$ and then all the parameters can be expressed
in terms of two unknowns; $m_{31}$ and $m_{22}$.  It is illustrative to consider
the limiting case of $t_{23}=1$ given the fact that the atmospheric neutrino
mixing is the best measured quantity; $t_{23}^2=1^{-0.3}_{+0.3}$.
Note in this case that we have the two solutions $p=0$ or $p=1$ along with the following
relations;
\bea
m_{32}=-m_{22} \quad\mbox{and}\quad m^2_{12}=2 {t^2_{12}}{ m^2_{22}}\,.
\eea
Equating \eq{eq:t23_13sec} and \eq{eq:t23_23sec} we find
\bea
 \frac{M_1}{M_2}=-\frac{m^2_{31}}{m^2_{22} }\, k \quad\mbox{with}\quad
k \equiv \frac{1}{2}[(1-p) - (1+p) c^2_{12}{\bf r} e^{2i\sigma}], \label{eq:fracM1M2}
 \eea
which leads us from \eq{eq:M2direct} to
\bea
 M_1=\frac{-2 m_{31}^2}{c^2_{12}{\bf r} m_{\nu_3}}\, k \,.
\eea
Now we find that \eq{eq:fracM1M2} is indeed compatible with \eq{eq:M1M2_23} for $p=1$.\\

\noindent{\bf{Form of the matrix} $\mathbf{m^\nu_{LR}}$}

Allowing a deviation from the limiting case of $t_{23}=1$, we can get a more
general values for the parameters. In any case,
we note that given $m_{22}$ we can determine the elements $m_{32}$
and $m_{12}$ through low energy observables; $t_{12}$ and
$t_{23}$ as in \eq{eq:tn_12_23}, and hence fix the scale of $M_2$ by using
$m_{\nu_3}$ and ${\bf r}$. Although we cannot fix the values of $m_{21}$
and $m_{31}$ independently, we can fix their ratio $p$ and satisfy
all experimental constraints. Then we can see how the hierarchy
of $M_1$ and $M_2$ depends on the ratio of $m^2_{22}$ and $m^2_{31}$.

 That is, we determine the following structure of $m^\nu_{LR}$
\bea
m^\nu_{LR}=\left[
\begin{array}{ccc}
0        & \frac{{\tn}_{12}}{{\cn}_{23}}m_{22} & x_1 \\
pm_{31}  & m_{22}                              & x_2\\
m_{31}   & -t_{23} m_{22}                      & x_3\end{array}
\right], \label{eq:dirac_mass}
 \eea
where the elements of the
third column cannot be determined due to the limit
$1/M_3\rightarrow 0$ we have taken, and we can see that all the
entries in the second column are of comparable order.
As we mentioned before,
the motivation for having $(m^\nu_{LR})_{11}\rightarrow 0$ was linked more closely to
the symmetric fits, which implies $m_{12} \sim m_{21}$.  From \eq{eq:M1M2_23}, we can write
\bea
\frac{M_1}{M_2}  \approx \frac{m_{21}^2}{m_{12}^2} {\bf r} s_{12}^2 \frac{1+p^2}{p^2},
\eea
which leads us to a conclusion of $M_1/M_2 \sim 0.2$ with $p=1$.
In the following, we will give more precise determination of the right-handed neutrino masses
further considering the $SO(10)$ structure of $m^\nu_{LR}$ in \eq{eq:dirac_mass}.

%
\subsubsection{${\mathbf{SO(10)}}$ and the scale of ${\mathbf{M}}_1$ \label{sec:so10realiz}}
%
%
%
We started our discussion by considering an underlying $SO(10)$
framework with the mass matrix relation (2.7).  Then,
it has been realized \cite{Ross:2002fb} that at least  one
non-renormalizable operator should be included in the Yukawa
sector to get a successful fit to the quark and charged-lepton sectors.
Generically the representations
which break the $SO(10)$ down to  $SU(5)$ or $SU(4)_C\times
SU(2)_R\times SU(2)_L$ and then further down to $SU(3)_C\times
SU(2)_L\times U(1)_Y$ do not give rise to the appropriate
structures of fermion masses, and thus we need to add at least
one non-renormalizable operator. As we mentioned,
 the non-renormalizable operator involving the
${\mathbf{45}}$ representation is quite useful since its vev can
be aligned in the $B-L+k T_{R3}$ direction for $k\in {\mathbf{Z}}$
whose  value is different for every species of fermions. As we
discussed in Section \ref{sec:gutconsid}, fits to the quark masses
\cite{Roberts:2001zy}-\cite{Ross:2004qn}, give a structure of the
up-sector of the form
\bea
Y^u\sim \left[
\begin{array}{ccc}
u & v & w \\
v & x & y \\
w & y & z
\end{array}
\right]\!\!,\quad u\ll v,\quad u< w \lsim v,\quad v < x,\quad y \lsim x \ll z \sim O(1),\label{eq:sym_y11_text}
\eea
where the exact symmetry of the matrix is not necessary  but the
orders of magnitude in the elements $Y^u_{ij}$ and $Y^u_{ji}$ have
to be the same. In the previous Section \ref{sec:limits130}, we
determined the possible form of $m^\nu_{LR}$ which can be
naturally understood in the context of the $SO(10)$ models when
$m_{31} \sim m_{22}$. There are two forms of Yukawa matrices for
the Dirac neutrinos that have been exploited in the literature:
\bea
\! Y^\nu&=&\left[
\begin{array}{ccc}
u^\prime       &  v^\prime_{12} & w^\prime_{13}\\
v^\prime_{21}  &  x^\prime      & y^\prime_{23}\\
w^\prime_{31}  &  y^\prime_{32} & z^\prime
\end{array}
\right]\!\!
\rightarrow \left\{
\begin{array}{c}
u^\prime \ll w^\prime_{13}\sim w^\prime_{31}\sim v^\prime_{12}\sim
v^\prime_{21},\quad
v^\prime_{12}\sim x^\prime\sim y^\prime_{32} \sim y^\prime_{23} \ll 1 \\
u^\prime \ll w^\prime_{13}\sim w^\prime_{31} \ll v^\prime_{12}\sim
v^\prime_{21},\quad v^\prime_{12}\ll x^\prime\sim y^\prime_{32} <
y^\prime_{23} \ll 1
\end{array}\right.\nn\\
\label{eq:form_of_ynu}
\eea
Each of these has been justified within the context of $SO(10)$ and
a particular flavour symmetry. The models \cite{Bando:2004hi} (BKOT),
\cite{Chen:2003zv} (CM) and \cite{Ross:2002fb} (RV) are examples of the
first form of $Y^\nu$ in \eq{eq:form_of_ynu}.  In
\cite{Ross:2002fb}, for instance, the behavior of  $x^\prime\sim w^\prime\sim
v^\prime$, which is different from   $v\ll
y\lsim x$ in the up-quark sector, was explained by the coupling of the elements $(2,2)$ and $(2,3)$
to a $\mathbf{45}$ representation whose vev is proportional to
$B-L + 2 T_{R,3}$. This vev is different from zero for up-quarks
while it vanishes for neutrinos and thus one is forced to take into account
the next leading contribution which has to be of the same
order for the entries $(1,2)$ and $(1,3)$  leading to a large mixing
solution. The model of \cite{Dermisek:2006dc} (DHR) is an example of the
second case. It has also a coupling of a $\mathbf{45}$
representation in the $(2,3)$ sectors. But the difference with
respect to the up-quark Yukawa couplings is given by the vev breaking the
flavour symmetry $D_3\times U(1) \times Z_2\times Z_3$ and
together with the other orders of magnitude in $Y^\nu$ a large mixing is achieved.
Now, \eq{eq:gst} fixes the order of $\epsilon_u$ as follows;
\bea
\epsilon_u\equiv\frac{v}{x}\approx \sqrt{\frac{m_u}{m_c}}\sim(3,6) \times 10^{-2},\quad
\frac{x}{z}\approx \frac{m_c}{m_t}\approx \epsilon^2_u,\quad z=1\ \Rightarrow v\approx \epsilon_u^3.
\eea
For the first case of \eq{eq:form_of_ynu}, $v^\prime$ is straightforwardly related to $v$:
\bea
v^\prime= v\sim \epsilon_u^3 \,.
\eea
On the other hand, for the second case of \eq{eq:form_of_ynu}, $v^\prime$ is also related
  to the up-sector but due to the structure of $D_3\times U(1) \times Z_2\times Z_3$  it is given by
\bea
v^\prime\sim \epsilon_u^2.
\eea
Once the scale of $m^2_{12}$ is fixed, we can determine the
order of magnitude of $M_2$ with $y^\nu_{12}=$ $c (5\times
10^{-2})^3$,
  \bea
M_2\approx\frac{y^2_{12}v^2 \sin^2\beta/2}{m^\nu_{ee}}
=c^2 \sin^2\beta \,(0.5,2.4)\times 10^{8} {\mathrm{GeV}},
\label{eq:estimM_2}
 \eea
where $c$ is an $O(1)$ number and  $m^\nu_{ee}$ is taken from
\eq{eq:numericmnu}  considering the normal hierarchical spectrum
of low energy neutrinos, ${\bf t}\ll {\bf r}$, and the
uncertainties in \eq{eq:oscill_inp} are taken into account. When ${\bf
r} \sim {\bf t}$, non-trivial phases can make $m^\nu_{ee}$ very
small and  the above estimation of $M_2$ has to be considered just
as a lower bound.
For \eq{eq:estimM_2} we have $a\ll 1$ and $b\ll 1$, and  hence this estimation
can be applied to  the BKOT, CM and RM models. For the model DHR
\cite{Dermisek:2006dc} having
$y_{12}=6.27 \times 10^{-3}=O(\epsilon_u^2)$, one gets much larger value:
$
M_2\approx (0.3,1.1) \times 10^{12} {\mathrm{GeV}}$
which agrees with the value quoted in Table \ref{tbl:so10mods}.
For the explicit case with $p=1$ analyzed in Section
\ref{sec:limits130}, the ratio of $M_1/M_2$ becomes
\bea
\frac{M_1}{M_2}=\frac{m^2_{31}}{m^2_{22}}(0.089,0.19),
\eea
Thus, we get the ranges of $M_1\approx (0.0045,0.46)\times10^8$ GeV.
%
%
%

%
%
\section{Baryogenesis through Leptogenesis}
%
%
\subsection{Thermal Leptogenesis}

The CP violating asymmetry generated in the decay of heavy right
handed neutrino into a Higgs boson and a left-handed lepton and
its CP conjugated channel, and its supersymmetric counterpart are
\bea \epsilon_{N_i}=\frac{\Gamma_{N_{i},l}-\Gamma_{N_{i},\bar{l}}}
{\Gamma_{N_{i},l}+\Gamma_{N_{i},\bar{l}}}, \quad \epsilon_{\tilde
N_i}=\frac{\Gamma_{\tilde N_{i},l}-\Gamma_{\tilde N_{i},\bar{l}}}
{\Gamma_{\tilde N_{i},l}+\Gamma_{\tilde N_{i},\bar{l}}} \eea
where $\Gamma_{N_{i}, l} \equiv \Sigma_{\alpha,\beta} \Gamma
(N_{i}\ \rightarrow\  l^\alpha H^\beta_d)$ and
$\Gamma_{N_{i},\bar{l}} \equiv \Sigma_{\alpha,\beta} \Gamma
(N_{R_i}\rightarrow \bar{l}^\alpha \bar{H}^\beta_d)$ are the $N_i$ decay rates
into  $l$ and $\bar{l}$ respectively.  For the right-handed
sneutrino $\tilde{N}_i$ decay rates, one has the final states with
the  lepton $l$ and slepton $\tilde{l}$.
The $B-L$ asymmetry generated by the right handed (s)neutrino
decays is given by
\bea Y_{B-L}&=&-\eta_i[\epsilon_{N_i} Y^{\mathrm{eq}}_{N_i} +
\epsilon_{\tilde{N}_i}Y^{\mathrm{eq}}_{\tilde N_i}] = {79\over 28}
Y_B
\label{eq:YB_fYBmL} \eea
where $Y_B$ is the resulting  baryon asymmetry converted from the
$Y_{B-L}$ by the electroweak sphaleron processes, and $\eta_i$ is
the efficiency factor that measures the number density of
$N_i/\tilde{N}_i$ decays at low temperature $T \ll M_i$ and
$Y^{\mathrm{eq}}=Y^{\mathrm{eq}}(T \gg M_i)=135 \zeta(3)/(4\pi^2
g_\star)$.  For  $g_\star=225$ in the supersymmetric model, one gets
$Y^{\mathrm{eq}}_{N_i}=1.9 \times 10^{-3}$. Since $N_i$ and
$\tilde{N}_i$ give the same contributions in the supersymmetric
limit; $\Gamma_{N_i}=\Gamma_{\tilde{N}_i}\equiv \Gamma_i$ and
$\epsilon_{N_i}=\epsilon_{\tilde N_i}\equiv \epsilon_i$,  one can
express the final baryon asymmetry as \bea Y_B =
1.3\times10^{-3}\, \eta_i\, \epsilon_{i} \eea where the
observation requires $Y_B \approx 10^{-10}$.
We recall that the decay rate for the right-handed (s)neutrino is
\bea \Gamma_i = {\Gamma_{N_{i},l}+\Gamma_{N_{i},\bar{l}}} =
\frac{(Y^{D\dagger}_\nu Y^D_\nu)_{ii}M_{N_i}}{4\pi}.
\label{eq:decay_rate} \eea
Then the CP asymmetry can be expressed by
\bea
\epsilon_i = \frac{1}{8\pi}
 \sum_{j\neq 1}\frac{{\mathrm{Im}}\left[ (Y^{D\dagger}_\nu Y^D_\nu)^2_{ji} \right]}
 {[Y^{D\dagger}_\nu Y^D_\nu]_{ii}}f\left(\frac{M^2_{j}}{M^2_{i}}\right),
\label{eq:th_cp_asym} \eea
where,  whenever the hierarchy $M^2_{N_j}/M^2_{N_i}=x$ is good
enough, the function $f(x)$ is just $f(x)\sim
-\frac{3}{2\sqrt{x}}$.

In our case, the right-handed neutrinos are charged under the
SO(10) gauge group and thus are in thermal equilibrium at high
temperature. The efficiency factor $\eta_i$ can be calculated given the value of
$K_i\equiv \Gamma_i/H(T=M_i)$;
\bea K_i=\frac{\tilde{m}_i}{m_*} = \frac{\tilde{m}_i}{1.6 \times
10^{-3} \mbox{ eV} }
\eea
where the effective neutrino mass $\tilde{m}_i$ is defined by
$\tilde{m}_i = 4\pi \Gamma_i v^2 / M_i^2$. If $K_i \lsim 1$, the efficiency
reached its maximum $\eta_i=1$. However, when $K_i \gg1$ as is the case in most
$SO(10)$ models, the
inverse decay remains effective for $T< M_i$ and its decoupling
temperature $z_i\equiv M_i/T_i$ is given by \cite{Buch}
 \bea
{K_i\over4} z_i^3 e^{-z_i} \sqrt{1+{\pi\over2} z_i} \simeq z_i -1
 \eea
and the corresponding efficiency factor can be well approximately
by the simple form \bea \eta_i \simeq {2\over z_i K_i}
\left(1-e^{-z_i K_i/2}\right). \eea

It is important for our case to notice that the lepton asymmetry
along the electron direction  generated by the second lightest
right-handed neutrino $N_2$ is not washed out by the lightest
right-handed neutrino $N_1$ as we have $y_{11}\to 0$.  Therefore,
let us consider the possibility of a successful leptogenesis
either from $N_1$ or $N_2$ whose effective neutrino masses are
 \bea
\tilde m_1&=&\frac{v^2}{M_{1}} \sin\beta^2 |y_{31}|^2[1+p^2]
\approx m_{\nu_3} \approx  0.05 {\mathrm{eV}} \,, \nn\\
\tilde m_2&=&\frac{v^2}{M_{2}} \sin\beta^2 |y_{22}|^2 \frac{1+
\tn_{12}^2}{\cn_{23}^2} \approx m_{\nu_2} \approx 0.009
{\mathrm{eV}} \,.
 \eea
Thus we get
\bea (K_i, z_i)= \cases{ (31.3, 7.51) \quad\mbox{for}\quad i=1\cr
                         (6.26, 4.96) \quad\mbox{for}\quad i=2 }
\eea leading to
 \bea
 \eta_i = \cases{ 8.5\times10^{-3}
\quad\mbox{for}\quad i=1\cr
                         6.4\times10^{-2} \quad\mbox{for}\quad i=2 }
 \eea
 On the other hand, one can readily check that  we have
 \bea
 O(\epsilon_{1,2})\sim
\frac{3}{16 \pi} y^2_{ij} \frac{M_1}{M_2}\lsim 10^{-10}
 \eea
putting the numerical values determined by the procedure of
Section \ref{sec:casemdeq0}. This is too small to produce the
required amount of baryon asymmetry, $10^{-10}$, as expected from
a general discussion with hierarchical neutrino mass spectrum
\cite{Davidson:2002qv}.

%
%
\subsection{Soft Leptogenesis}
It has been pointed out \cite{D'Ambrosio:2003wy} that the $B$-term
soft supersymmetry breaking of the right-handed sneutrino provides
an additional source of lepton number and CP violation, where the
relevant couplings are given by
\bea
-{\mathcal{L}_{\rm{soft}}}&=&
    (m^2_{\tilde{N}})^j_i \tilde{N}^{* i}_R \tilde{N}_{Rj}
+ (a^{ije}\tilde{l}_{Lj} \tilde{e}^{*}_{Ri} h_d
+ a^{ijN}\tilde{l}_{Lj} \tilde{N}^{*}_{Ri} h_u + {\mathrm{h.c}})\nn\\
&+& \left(\frac{1}{2}(b_{N})^i_j \tilde{N}^{* i}_R \tilde{N}^*_{Rj}+ {\mathrm{h.c.}}\right).
\label{eq:Lsoft}
\eea
 The effects of $b_{N}\equiv B M_R$ terms are usually ignored because they are assumed to be
highly suppressed by the difference in scales of the typical
supersymmetric masses, $10^3$ GeV, with respect to the masses of
the singlet neutrinos, $M_R \geq 10^7$ GeV. It turns out that there is a region in the
parameter
 space of  $B$ and $M_R$ compatible with models for which the masses of right-handed neutrinos are as low as
 $M_R\sim (10^7-10^8)$ GeV.
 The non-vanishing value of the generated lepton asymmetry is a pure thermal effect
 since at $T=0$ the generated lepton asymmetry in leptons cancels the one in sleptons:
\bea \epsilon_{\tilde N_i\rightarrow \tilde l
H_d}=-\epsilon_{\tilde N_i\rightarrow l \tilde{H}} =\frac{4
\Gamma_{\tilde N_i} B }{4 B^2 +\Gamma^2_{\tilde
N_i}}\frac{[\mathrm{Im}A]}{M_i}\,, \eea
where  $A=a_N/Y^\nu$.  At finite temperature
$T\neq 0$, the difference between the fermion and boson statistics
yields non-vanishing lepton and CP asymmetry of the form
$\epsilon_i(z)= \epsilon_{\tilde{N}_i\rightarrow \tilde l H_d}
\delta_{BF}(z)$ where $\delta_{BF}(z)$ can be approximated for
$z\gg1$ by the analytic function of $\delta_{BF}(z) \equiv
2\sqrt{2} K_1(\sqrt{2}z)/K_1(z)$ with $K_1(z)$ is the modified
Bessel function of the first kind \cite{scopel}.

Thus, in the soft
leptogenesis, one gets the reduced efficiency defined by
 \bea
\tilde{\eta}_i \approx 2\sqrt{2} {K_1(\sqrt{2}z_i)\over K_1(z_i)}
\times \eta_i
 \eea
where $z_i$ is the decoupling temperature of the inverse
decay calculated before. For our case with $z_i=7.51$
and  $4.96$, we get $\delta_{BF}(z_i) = 0.105$ and $0.3$ and thus
\bea
 \tilde{\eta}_i \approx \cases{8.9\times10^{-4}
\quad\mbox{for}\quad i=1\cr  1.9\times10^{-2}\quad\mbox{for}\quad
i=2 }.
 \eea
Therefore, we require $\epsilon_{\tilde{N}_i\rightarrow \tilde l
H_d} \approx 8.6\times10^{-5}$ and $4.0\times10^{-6}$
correspondingly for a successful leptogenesis.
When $\Gamma_i < B$, we get
  \bea
 \epsilon_{\tilde{N}_i\rightarrow
\tilde l H_d} \approx {\Gamma_i \over B} { {\rm Im}[A]\over  M_i}
= {\tilde{m}_i M_i \over 4\pi v^2} {{\rm Im}[A]\over B}.
 \eea
Now one can find that the lightest right-handed sneutrino $\tilde
N_1$ cannot produce enough lepton asymmetry  due to a strong
wash-out suppression. However, in the case of $\tilde N_2$ with
$M_2=10^8$ GeV, the desired value of
$\epsilon_{\tilde{N}_2\rightarrow \tilde l H_d} \approx
4.0\times10^{-6}$ is found to be achieved for the hierarchical
choice of soft parameters; $\rm{Im}[A]\approx 1.7$ TeV and $B=1$
GeV.
Note that the leptogenesis scale $\sim 10^8$ GeV can be marginally allowed
in view of abundant unstable gravitinos which decay late and upset the standard prediction of the big-bang neucleosynthesis \cite{bbn1}.  Recent analyses showed that the upper bound on the reheat temperature is $T_R=2\times10^6-3\times10^8$ GeV for the typical gravitino mass range of $m_{3/2}=10^2-10^3$ GeV, assuming the hadronic branching ratio of the gravitino decay is $10^{-3}$.  We also remark that the bound on the reheat temperature can be loosened if the gravitino is stable and forms dark matter.  In this case,
the next lightest supersymmetric particle needs to be a stau and the reheat
temperature up to $T_R=10^{10}$ GeV can be acceptable \cite{bbn2}.

%
%
\section{Supergravity description of a small $B$ term}
%
%
It has been pointed out  that the smallness of the $B$ term may
arise if we have for example a dynamical mechanism that sets $B=0$ at the leading order by
 arranging a specific form of the superpotential Ref.~\cite{Yamaguchi:2002zy}.
However, we find it difficult to achieve such a mechanism without introducing a fine
tuning of parameters in the general supergravity context. In this scenario,
the other ingredient to produce a small $B$ term of the order $m^2_{3/2}/M_{N_1}$
it may be through a term $\int d \theta^4 X^\dagger X N_1 N_1$ as in \cite{Giudice:1988yz}-\cite{Kim:1994eu}.

To illustrate difficulties in a dynamical set up of $B=0$ in supergravity,
let us consider the superpotential suggested in Ref.~\cite{Yamaguchi:2002zy}
 \bea
W= \mu(\Phi_i) N  N + A f(\Phi_i) X + W^\prime,
\label{eq:genW}
 \eea
where $\Phi_i$ can be an observable field such as an multiplet of
SO(10) (that is, {\bf 126} as $N$ can be {\bf 16}) or any field in
the hidden sector.   The minimization condition of the scalar potential
$V=e^{K}[K^{i\bar{j}} F_iF_{\bar{j}}-3|W|^2]$ reads $V_l=0$ where
 \bea
 V_l=e^K\left[
K^{i\bar{j}}_{l}F_i F_{\bar{j}}+K^{i\bar{j}}
\left[F_i(\bar{W}K_{\bar{j} l})+  F_{\bar{j}}(W_{il} +K_i W_l+
K_{il}W )  \right]-3W_l\bar{W}\right]+K_l V \label{eq:Vl} \eea
 is  the derivative of the potential $V$ with respect to a field
$l$. Here we have defined $F_l=W_l+K_l W$.
The
$b$ term coming from the scalar potential is given by
\bea b=e^K\left[ K^{i\bar{j}}F_{\bar{j}}(\mu_i +K_i \mu )-3\mu
\bar{W} + \underline{ 2 \mu \bar{W}} \right] \label{eq:btermex}
\eea
where the last term comes from
$|F_N|^2= |2 \mu N + K_N W|^2$ with
the minimal kinetic term for $N$; $K_N= \bar{N}$.
{On the other hand the minimization condition for $X$ is
\bea
 0=
K^{i\bar{j}} \left[F_i(\bar{W}K_{\bar{j} X})+ F_{\bar{j}}(W_{iX}
+K_i W_X+ K_{iX}W )  \right]-3W_X\bar{W}, \label{eq:VX}
 \eea
assuming $K^{i\bar{j}}_X=0$ and $V=0$ at the minimum. The indices
$i$ and $j$ in the above equation contain $\Phi_i$ and $X$.
Assuming there is no mixing term between them in $K$ (that is,
$K_{\Phi_i \bar{X}}=0$ and $K_{\Phi_i X}=0$), we separate the $X$
index to write
\bea
  0=K^{i\bar{j}} \left[F_{\bar{j}}(W_{iX} +K_i
W_X ) \right]-3W_X\bar{W} \nonumber\\+ K^{X\bar{X}} \left[ F_X
\bar{W} K_{\bar{X}X} + F_{\bar{X}}(W_{XX} + K_X W_X + K_{XX}
W)\right]. \label{eq:VXandb0}
  \eea
  Thus we need to arrange the
second line of \eq{eq:VXandb0} and the $W_{iX}$ contribution to
sum up to $2 \mu \bar{W}$ to cancel the above $b$ term in
\eq{eq:btermex}. We find that, with an specific form of a K\"ahler potential,
one can achieve such a condition which however requires a fine tuning of
the parameters involved and does not have a real theoretical
justification.

 The simplest way to arrange the condition of $B=0$ is to rely on the
no-scale supergravity models as in Ref.~\cite{chun}. For this, let
us take the hidden sector field $\phi$ with a  K\"alher potential
$$ K= -3 \log(\phi+\phi^*),$$ and its Yukawa coupling to matter fields
 \bea
Y_{10,120}(\phi) = e^{-c \phi},\qquad Y_{126}=const.
 \eea
This can be a consequence of  an $U(1)$ symmetry under which
$\phi$ transforms like $\phi\to \phi + i \alpha$, and then ${\bf
16.16.10}$ and ${\bf 16.16.120}$ are charged  but ${\bf
16.16.\overline{126}}$ is not.  The $A$ terms associated to the Yukawa couplings are
given by \cite{Brignole}
 \bea
 A Y = -m_{3/2} (\phi+\phi^*)
\partial_\phi Y.
 \eea
Therefore, we obtain $A_{10,120}\sim m_{3/2}$ and $B=A_{126}=0$ at
the GUT scale.

On the other hand the smallness of the $B$ term needed for a successful soft
leptogenesis could follow simply from a tuning among various
supersymmetry breaking terms, which is technically
natural if it is stable under sub-leading corrections.
It is amusing to realize that the $B$ term  receives an important radiative correction
due to gauge interactions of the right-handed (s)neutrinos.   In the context of
$SO(10)$, the right-handed (s)neutrinos have a coupling to
a heavy gauge boson $X$ and the corresponding gaugino $\tilde{X}$
which also obtains a supersymmetry breaking mass $m_{1/2}$.
Specifically, the gauge coupling $\tilde{N}$--$N$--$\tilde{X}$
leads to the one-loop correction to the $B$ term of $\tilde{N}$
which is given by
 \bea
 B M \approx {\alpha\over 4\pi} m_{1/2} M \log{M_X \over M}
 \eea
where $M_X$ is the mass scale of the heavy gauge boson $X$ or the
$B-L$ symmetry breaking, for instance. Now, putting $\alpha=1/30$,
$M=10^8$ GeV and $M_X=10^{10}$ GeV, we find $B\approx 10^{-2}
m_{1/2}$ which gives us the required value of $B\approx 1$ GeV for $m_{1/2}=100$ GeV.

%
%
%
\section{Conclusions}

The motivation of our work was to understand why in many GUT models which describe successfully the right values of fermion masses and mixings (\cite{Ross:2002fb},
\cite{Chen:2004xy}, \cite{Bando:2004hi}, \cite{Dermisek:2006dc}),
it is not possible to achieve the observed baryon asymmetry through
the decay of heavy right-handed neutrinos. Apart from the strong hierarchy
in the neutrino Yukawa couplings $Y^\nu$, two factors are important: first linking
$m^u$ and $m^\nu_{LR}$ (in the simplest case they could be the same),
which is a general $SO(10)$ GUT relation, and then having the special
feature of $Y^\nu_{11}\to 0$. 
Starting from these conditions we reconstructed the general structure of $Y^\nu$
and the mass scales of two right-handed neutrinos which are
compatible with the neutrino data as well as the GUT relations
enforced by a certain flavor symmetry in the decoupling limit of
the heaviest right-handed neutrinos.

 Our analysis shows that the
neutrino couplings associated with two light right-handed
neutrinos are determined by $Y^\nu_{i1, j2} \sim \epsilon_u^3 \sim
10^{-4}$ while the right-handed neutrino masses are of order
$(10^7-10^8)$ GeV.  Conventionally, such a parameter region is far
away from a successful leptogenesis unless a certain fine-tuning
is arranged between two light right-handed neutrinos to resonantly
enhance the resulting lepton asymmetry.  However, the soft
leptogenesis arising from the CP phases of $A$ and $B$
supersymmetry breaking soft parameters can work consistently with
our picture although our parameters are in the strong wash-out
regime of the lepton asymmetry.  The basic ingredients for this to
occur are (i) $Y^\nu_{11}\to 0$ and (ii) a resonance condition of
$B\sim \Gamma$.  The first property protects  the electron
asymmetry which is generated by the second lightest right-handed
sneutrino whose wash-out factor is favorably smaller.
Interestingly the resonance condition requiring $B\sim 1$ GeV can
be a consequence of the gauge one-loop correction involving the
coupling of the right-handed sneutrino to the heavy GUT gaugino.
For the vanishing condition of the tree-level $B$ term, one may
invoke no-scale supergravity, as it is difficult to achieve a dynamical realization
of $B=0$ by arranging a specific form of the superpotential and K\"ahler potential,
 which requires to introduce a fine tuning of parameters in the general supergravity context.

\acknowledgments
L. Velasco-Sevilla would like to thank M. Peloso, K. Olive, A.
Vainshtein,  K. Kadota and S. Raby for helpful discussions. L.
Velasco-Sevilla is supported in party by the DOE grant
DE-FG02-94ER.  We thank I.Masina for useful comments regarding alternatives of possible Yukawa matrices compatible with thermal leptogenesis.

\appendix

\section{Numerical form of $m_{\nu}$\label{sec:appn_um_mnu}}

The order of magnitudes in $m_\nu$ can be illustrated in the case
$m_{\nu_3}\gg m_{\nu_2}>m_{\nu_1}$, for which we determine
numerically the approximate values of $m_\nu$ up to the value of
${\bf t}< {\bf r}$,  \eq{eq:assumhierch}, and the limit
$s^2_{13}\leq 0.041$:
\bea
\frac{(m_\nu)_{ee}}{m_{\nu_3}}&=& 0.0542 + s_{13}^2 + 0.719 {\mathbf{t}}\nn \\
\frac{(m_\nu)_{e\mu}}{m_{\nu_3}}&=&0.102(0.600 - 0.375\ s_{13}) + \frac{s_{13}}{\sqrt{2}} - 0.848(0.375 + 0.600  s_{13})\ {\mathbf{t}}\nn\\
\frac{(m_\nu)_{e\tau} }{m_{\nu_3}}&=& - 0.102 (0.600 + 0.375\ s_{13}) + \frac{s_{13} }{\sqrt{2}} + 0.848 (0.375 - 0.600 s_{13}) {\mathbf{t}} \nn \\
\frac{(m_\nu)_{\mu\mu}}{m_{\nu_3}}&=& \frac{1}{2} + 0.193(0.600 - 0.376\ s_{13})^2 - (0.375 - 0.600\ s_{13})^2 {\mathbf{t}}\nn \\
\frac{(m_\nu)_{\mu\tau}}{m_{\nu_3}}&=& \frac{1}{2} - 0.193(0.600 - 0.376\ s_{13})(0.600 - 0.376 s_{13}) \nn\\
&&- (0.375 + 0.600\ s_{13}) (0.375 - 0.600 \ s_{13}) {\mathbf{t}}\nn \\
\frac{(m_\nu)_{\tau\tau}}{m_{\nu_3}}&=& \frac{1}{2} - 0.193 (0.600 + 0.375 s_{13})^2 + (0.375 -
0.600 s_{13})^2 {\mathbf{t}}.
\label{eq:numericmnu}
\eea

\section{Fit to a particular form of symmetric Yukawa matrices \label{app:fit}}

\subsection{Assumptions}

A fit of textures for up and down Yukawa matrices of the form
\eq{eq:sym_y11_text} with $Y_{11}$ negligible can be made
compatible with the experimental information, under the following
theoretical assumptions:

\noindent{\bf(a)}  Yukawa matrices are the only source of CP
violation.

\noindent{\bf(b)}  We have a supersymmetric scenario which
respects at low energy the constraints of unitarity of the CKM
matrix.

\noindent{\bf(c)}  The supersymmetric corrections to the ratios
$\frac{m_u}{m_d}$, $\frac{m_c}{m_s}$ and $\frac{m_s}{m_b}$ do not
exceed the percentage of error on those ratios as quoted in Table
\ref{tbl:input_par}.

Since we have assumed {\bf(a)} and  {\bf(b)},  we must make the
fits that test the unitarity of the CKM matrix in the Standard
Model, where all experimental information has been taken into
account, rather than to specific experiments.
  We use the classical fit \cite{ckmfitt_coll}, which takes into account
  the measurements of the following flavour violating processes:
\bea \frac{|V_{ub}|}{|V_{cb}|},\quad
\frac{|V_{td}|}{|V_{ts}|},\quad \epsilon_k,\quad \dmd, \quad
\dms,\quad {\mathrm{and}} \quad \sin\beta. \label{eq:DF2_const}
\eea These fits are a test of the unitarity of the CKM matrix in
the Standard Model.  That is, if all the experimental inputs in
\eq{eq:DF2_const} are in agreement with the unitary of the CKM
matrix, the statistical and systematic errors are under control on
those measurements and there is no sensitivity to physics beyond the
Standard Model in those processes then {\it after} the fit all
these {\it fitted} quantities will agree with their input
values at the $68\%$ confidence level (C.L.). If some of them
do not agree then there is an indication of either (i) the
departure of the unitarity of the CKM matrix, (ii) a large correction
from statistical or systematic errors in the experimental
measurements, or (iii)  a contribution from process beyond the
standard model to the constraints \eq{eq:DF2_const} at the level
of sensitivity at which the measurements and the analyses are
performed. The other part of our experimental inputs comes from
considering the following mass ratios
\bea
\frac{m_u}{m_d},\quad \frac{m_c}{m_s},\quad \frac{m_s}{m_b},
\eea
at the scale $M_Z$, as well as the chiral perturbation parameter
$Q$ \cite{pdg07}: \bea Q=\frac{m_s/m_d}{\sqrt{1-(m_s/m_d)^2}}.
\eea
We take into account the renormalization from the low scales,
at which the quark masses are measured or computed according to
experimental data, up to $M_Z$ as
\bea \eta_i \equiv \frac{m_i(M_Z)}{m_i(m_i)}  \ {\rm for} \;\; i =
c,\ b,\ t\; ;  \qquad  \eta_i \equiv \frac{m_i(M_Z)}{m_i(2 \ {\rm
GeV})}  \ {\rm for} \;\; i = u,\ d,\ s.\nn
\eea
 At two loops in QCD they can be estimated to be $\eta_c = 0.56$, $\eta_b = 0.69$, $\eta_t =
1.06$, $\eta_u$ $=\eta_d$  $=\eta_s= 0.65$.

The form of the Yukawa matrix \eq{eq:sym_y11_text} is thought as
being compatible with supersymmetric  $SO(10)$ models for which
$\tan\beta$ is large, ($\sim$ 40-50). In this case the
corrections to quark masses are not negligible, e.g., for the $b$
quark can be up to $20\%$ \cite{Pierce:1996zz}. Thus strictly
speaking we have
\bea m_{f_d} = \sqrt{2} M_W \frac{y_{f_d}}{g}\cos\beta_S
(1+\varepsilon_{f_d} \tan\beta_S),\quad m_{f_u} = \sqrt{2} M_W
\frac{y_{f_u}}{g}\sin\beta_S (1+\varepsilon_{f_u} \tan\beta_S),\nn
\eea where the parameters $\varepsilon_{f,u}$ depend on the
supersymmetric  particles, such as charginos, neutralinos and
gluinos, and $y_{f_i}$  are eigenvalues of the Yukawa matrices.
However we assume here that the ratios are not strongly affected
by those corrections:
\bea
\frac{m_u}{m_c}=\frac{1+\epsilon_{u}}{1+\epsilon_{c}}\frac{y_u}{y_c}\equiv \frac{1}{r_{uc}} \frac{y_u}{y_c},\quad r_{uc}\approx 1,
\label{eq:susy_corr_mq}
\eea
analogously for the other mass ratios considered.

Under the conditions of \eq{eq:sym_y11_text}  then we expect the
angles of the left diagonalization matrices to be given as
 \bea
s^u_{12}&=&\sqrt{\frac{y_u}{y_c}}\quad  \rightarrow \quad
\sqrt{\frac{m_u}{m_c}r_{uc}},\qquad
s^d_{12}\ =\ \sqrt{\frac{y_d}{y_s+y_d}}\quad   \rightarrow  \quad
\sqrt{\frac{\frac{m_d}{m_s} r_{KS} }{1+\frac{m_d}{m_s}r_{ds} }}  ,\nn\\
s^d_{23}&=&\sqrt{\frac{\frac{Y^d_{22}}{Y^d_{33}}-\frac{y_s}{y_b}}{2\frac{y_s}{y_b}+1}}\quad
\rightarrow \quad
\sqrt{\frac{\frac{Y^d_{22}}{Y^d_{33}}-\frac{m_s}{m_b}
r_{sb}}{2\frac{m_s}{m_b}r_{sb}+1}}, \label{eq:mix_angl_th},
 \eea
where $r_{ab}$ are defined as in  \eq{eq:susy_corr_mq}, and
$s^d_{12}$ now contains $m_d$ in the denominator, which is a
correction to the approximate formula $s^d_{12}=\sqrt{m_d/m_s}$.
The angle $s^d_{23}$ is obtained, of course, assuming that it is
small and extracted from the relation,
\bea
s^d_{23}=\frac{Y^d_{23}}{Y^d_{33}}\quad \rightarrow \quad \frac{y_s}{y_b}=
\frac{Y^d_{22}-Y^d_{33}s^{d 2}_{23}}{Y^d_{33} (1+ 2 s^{d 2}_{23})}. \nn
\eea
The rest of the diagonalization angles are subject to the following conditions:
\bea
s^u_{13} \ll s^u_{23} \ll s^u_{12},\qquad
s^u_{13} \lesssim s^d_{13} \ll s^d_{23},\qquad
s^u_{23} \ll s^d_{23}.
\label{eq:cond_diag_angles}
\eea
If we assume that the phases of the elements $Y^d_{23}$ and $Y^u_{23}$
vanish or are the same, we can describe then the angles defining
the CKM matrix in terms of two phases, $\phi_1$, $\phi_2$
\cite{Ross:2004qn}: \bea
s_{12}e^{i\varphi_{12}} & = & s^d_{12}-c^d_{12}s^u_{12}e^{i\phi_1}\nn\\
s_{13}e^{-i\delta}      & = & s^d_{13}e^{i(\phi_2-\varphi_{12})} -s^u_{12} s^d_{23} e^{i(\phi_1-\varphi_{12})}\nn\\
s_{23}                  & = & s^d_{23},
\label{eq:angles_st__forckm}
 \eea
where then $\phi_1$ and  $\phi_2$ will be given as combination of
the phases of the elements $Y^f_{ij}$, except for $(i,j)=(2,3)$
and for $Y^u_{11}$ that we are neglecting.
Hence the CKM elements are expressed in terms of
 \bea
\left|{V_{ub}}\right| &=& |s^u_{12}s^d_{23}-s^d_{13}e^{i(\phi_2-\phi_1)}|\nn,\\
|V_{cb}|              &=& |s^d_{23}|, \nn\\
\left|\frac{V_{td}}{V_{ts}}\right| &=&
\left|\frac{s^d_{12}-c^d_{12}s^u_{12}e^{-i\phi_1}}{c^d_{12}}
-\frac{1}{s^d_{23}}\left[s^d_{13}e^{-i \phi_2}-s^u_{12}s^d_{23}e^{-i\phi_1} \right] \right| =|t^d_{12}-\frac{s^d_{13}}{s^d_{23}}e^{-i\phi_2}|,\nn,\\
|V_{us}|              &=& |s^d_{12}-s^u_{12}c^d_{12}e^{i\phi_1}|,\nn\\
{\rm Im} J            &=& (s^d_{23})^2c^d_{12}\left[ s^d_{12}
s^u_{12} \sin\phi_1 + c^d_{12} s^u_{12} K \sin(\phi_1-\phi_2) -
s^d_{12}K\sin\phi_2\right],
 \label{eq:model_ckm_elts}
 \eea
where $K\equiv {s^d_{13}}/{s^d_{23}}$.
 In \eq{eq:model_ckm_elts}, the quantities
that are not given by the form of the matrix \eq{eq:sym_y11_text}
or by the conditions \eq{eq:cond_diag_angles}, are
 \bea
s^d_{13},\quad \phi_1\quad {\mathrm{and}}\quad \phi_2.
\eea
 Assuming
$ s^d_{13} \ll s^d_{23}$ and given \eq{eq:mix_angl_th}, we have
$s^d_{23} < s^d_{12}$. Also from the dependence of $V_{td}/V_{ts}$
on $s^d_{13}$ and $s^d_{23}$ and from the fact that we are fitting
to $s^d_{23}$ directly because we are effectively making
$V_{cb}=s^d_{23}$, we can  fit to $K$ and the phases $\phi_1$ and
$\phi_2$. Here we choose to fit also $\phi_1$ to check the level
of compatibility of having $\phi_1=\pi/2$.

\subsection{Method of the fit}

Note that we do not know all the entries in \eq{eq:model_ckm_elts}
and we have 4 CKM parameters and 3 mass ratios from which we can
fit.  Thus, instead of using a $\chi$ method, we use a Bayesian
approach where we can obtain the combined probability distribution
for $K$, $\phi_2$ and $\phi_1$, which is identified with the
likelihood;
\bea {\cal{L}} (K, \phi_1, \phi_2) &\propto& \int \prod_{j=1,M}
f(\hat{c_j} | c_j(K, \phi_1, \phi_2, \{x_i\})) \times
\prod_{i=1,N} f_i(x_i) ~dx_i
~\times  f_0(K, \phi_1, \phi_2),\nn \\
\label{eqn:text_st:pdf-re}
 \eea
where $f(\hat{c_j} | c_j(K, \phi_1, \phi_2 \{x_i\}))$ is the
conditional probability density function (pdf) of the constraints
$c_j=$ ~ $|V_{ub}|$, $|V_{cb}|$,$V_{us}$, $|V_{td}|/|V_{ts}|$ and
${\rm{Im}}\{J\}$ given their dependence as functions of the
texture parameters: $s^d_{12}$, $s^u_{12}$, $c^d_{12}$,
$s^d_{13}$and $s^d_{23}$ as well as the parameters and
$x_i=\{m_u/m_d,\ m_c/m_s,\ m_s/m_b, Q \}$.

We have taken the values of the constraints  $c_j=$ ~ $|V_{ub}|$,
$|V_{cb}|$,$V_{us}$, $|V_{td}|/|V_{ts}|$ and ${\rm{Im}}\{J\}$ from
the most recent  CKM fitter results \cite{ckmf07} , listed in
Table \ref{tbl:input_par}. The values of the observables
\bea
\sin 2 \alpha, \quad \sin 2 \beta,\quad \sin 2 \gamma,
\eea
that we have to compare to our fit are then the values from the
same CKM fitter  results. Here
\bea
&& \alpha={\rm{Arg}}[-V_{31}V^*_{33}/(V_{11}V^*_{13})], \quad
\beta={\rm{Arg}}[-V_{21}V^*_{23}/(V_{31}V^*_{33})] \nn\\
&& \gamma={\rm{Arg}}[-V_{11}V^*_{13}/(V_{21}V^*_{23})],\nn
\eea
are the angles of the unitary triangle of the CKM matrix $V$.

For our fit then the fitted values of the parameter of texture
\eq{eq:sym_y11_text}  will be in agreement with the fits of the
CKM fitter of the unitary triangle, assuming that the
supersymmetric contributions are not relevant at the sensitivity
at which the parameters \eq{eq:DF2_const} are related to their SM
counterpart.

\subsection{Results and comments}

In Figure \ref{fig:2D_K_phi1_phi2}, we show the results for the 2D
probabilities of the parameter $K$ versus $\phi_2$ and $\phi_1$.
 Given the method used for this fit we expect at least a
$95\%$ C.L. compatibility with what we have fitted. If that is not
the case then obviously one of our theoretical assumptions should
be modified. However we do have a $68\%$ C.L. compatibility of all
of our input values with the output (fitted) information. We show
in Table \ref{tbl:input_par} the input values used for the fit. We
have chosen to use the results of the CKM fitter collaboration
\cite{ckmf07} (which provides the SM fits to the Particle
Data Group). In Table \ref{tbl:other_ckm_par} we show for
comparison the different values of the unitary angles. In Table
\ref{tbl:outputs} we show the results of our fit.

\begin{figure}[h]
~\quad
\begin{minipage}[t]{6.8cm}
\begin{center}
\includegraphics[width=1\textwidth]{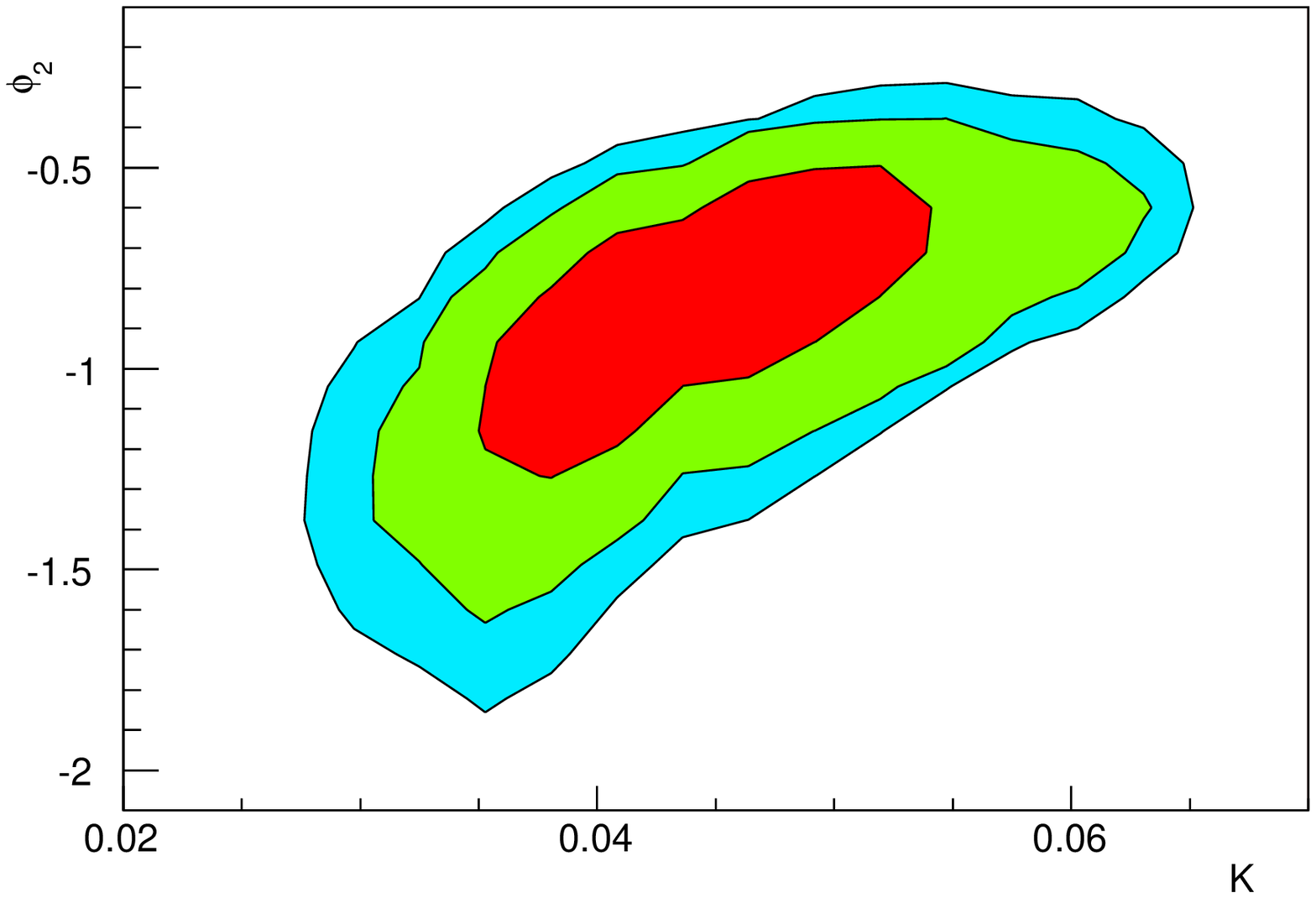}
\vspace*{-0.3cm}
(a)
\end{center}
\end{minipage}
~
\begin{minipage}[t]{6.8cm}
\begin{center}
\includegraphics[width=1\textwidth]{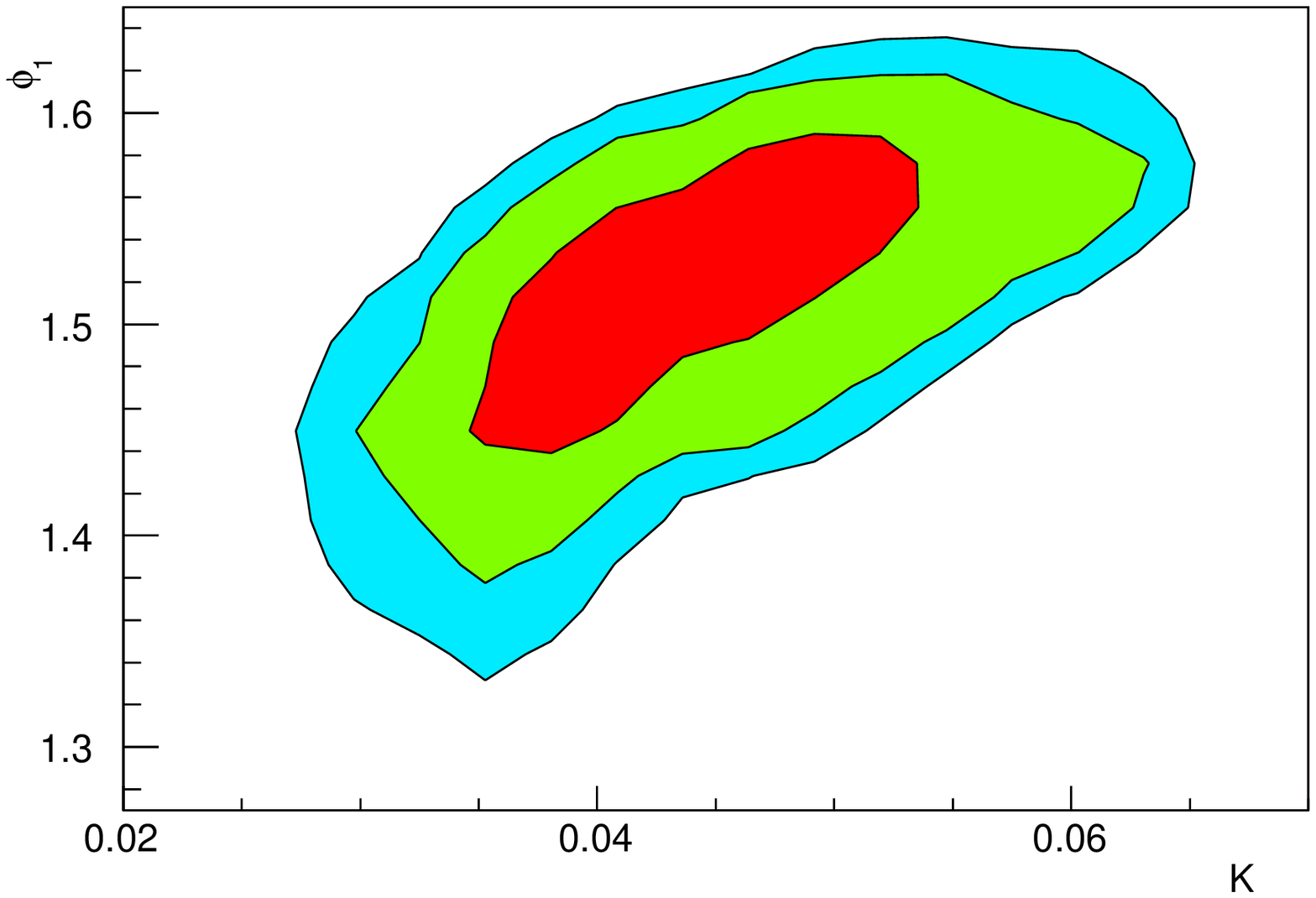}
(b)
\end{center}
\end{minipage}
\hfill \vspace*{-0.3cm} \caption{\small 2D probabilities (a) for K
and $\phi_1$  and (b) for K and $\phi_2$, the C.L. shown are at
68\%, 95\% and 99\%.} \label{fig:2D_K_phi1_phi2}
\end{figure}
\begin{table}
\begin{center}
\begin{tabular}{|l l l l l|}
\hline
\multicolumn{5}{c}{Input Values}\\
\hline
Constraints            & Value $\pm$ Gaussian errors      & Flat errors            &$\%$ of error & Referen.\\
\hline
&  & & &\\
$|V_{ub}|$       & $(3.683^{+0.106}_{-0.079} )\times 10^{-3}$   &                      &  & \cite{ckmf07} \\
$|V_{cb}|$       & $(41.61^{+0.62}_{-0.63})\times 10^{-3}$   &  & &  ``  \\
$|V_{td}|/|V_{ts}|$      & $0.2003^{+0.0146}_{-0.0059} $     &  &  &  `` \\
$|V_{us}|$       & $0.22715^{+0.00101}_{-0.00100}$   &  &  &  `` \\
${\rm{Im}}\{J\}$ & $(2.91^{+0.25}_{-0.14})\times 10^{-5}$  & &  & ``\\
Varied Parameters.   &                    &     & &  \\
$\frac{m_u}{m_d}$    & $0.553 \pm 0.043$  &     & $7.7\%$    & \\
$\frac{m_c}{m_s}$    & $11.3  \pm 2.8$    &     & $33.5\%$ & \\
$\frac{m_s}{m_b}$    & $0.0213\pm 0.006$  &     & $28\%$ & \\
$Q$                  & $22.7  \pm 0.8$    &     &  & \\
$\frac{y^d_{22}}{y^d_{33}}$  & $0.036$    &$\pm 0.014$ & &\\
\hline
\end{tabular}
\end{center}
\caption{\small Input values for constraints and varied parameters
which are also fitted.} \label{tbl:input_par}
\end{table}
\begin{table}
\begin{center}
\begin{tabular}{|l l l l l |}
\hline
\multicolumn{5}{c}{Other values from the CKM fitter}\\
\hline
Parameter            & CKM value   $\pm 1\sigma$ C.L. &  $\pm 2\sigma$ C.L.
& Direct exp. value  $\pm 1\sigma$ C.L. &  $\pm 2\sigma$ C.L.   \\
\hline
$\alpha$ & $99.0^{+4.0}_{-9.4}$    &   $^{+8.0}_{-17.9}$   & $92.6.0^{+10.7}_{-9.3}$
& $^{+27.1}_{-15.7}$  \\
$\beta$  & $22.03^{+0.72}_{-0.62}$ &   $^{+1.69}_{-1.27}$  & $21.23^{+1.03}_{-0.99}$
& $^{+2.09}_{-1.96}$ \\
$\gamma$ & $59.0^{+9.2}_{-3.7}$    &   $^{+18.0}_{-7.3}$   & $60.0^{+38}_{-24}$
& $^{+62}_{-39}$ \\
\hline
\end{tabular}
\end{center}
\caption{\small Relevant information from experiments and from the
CKM fitter \cite{ckmf07}. The later are included indirectly in the
fit because they are not used as constraints.}
\label{tbl:other_ckm_par}
\end{table}
\begin{table}[!h]
\begin{center}
\begin{tabular}{|l l |}
\hline
\multicolumn{2}{c}{Output Values}\\
\hline
Parameter            & Value $\pm$ errors     \\
\hline
           &  \\
$K$                         & $0.044^{+0.005}_{-0.007} $ \\
$\phi_1$                    & $1.530^{+0.050}_{-0.030} $ \\
$\phi_2$                    & $-0.877^{-0.162}_{+0.271}$ \\
$\frac{y^d_{23}}{y^d_{33}}$ & $0.023^{-0.065}_{+0.066}$ \\
$|V_{ub}|$       & $(3.75^{+0.096}_{-0.067} )\times 10^{-3}$ \\
$|V_{cb}|$       & $(42.25^{+0.58}_{-0.60})\times 10^{-3}$  \\
$|V_{td}|/|V_{ts}|$      & $0.2046^{+0.0096}_{-0.0057} $  \\
$|V_{us}|$       & $0.22751^{+0.00091}_{-0.00081}$  \\
${\rm{Im}}\{J\}$ & $(3.02^{+0.18}_{-0.11})\times 10^{-5}$\\
$\alpha$   & $(82.14^{+8.4}_{-8.5})^o$\\
$\beta$    & $(21.83^{+2.3}_{-1.83})^o$\\
$\gamma$   & $\approx \delta= (60.3^{+5.5}_{-5.3})^o$ \\
 \hline
\end{tabular}
\end{center}
\caption{\small Our output values} \label{tbl:outputs}
\end{table}

We can notice indeed that while the output angles $\beta$ and
$\gamma$ are fitted in great agreement with the inputs, the angle
$\alpha$ tends to be lower than the CKM fitter central value,
although compatible at $68\%$ C.L.
This tendency could be due to the fact that the unitary triangle
(UT) fit itself have shown consistently during the last 5 years a
difficulty in fitting the unitary condition itself: $\pi=\alpha +
\beta +\gamma$ (which is used in such fits with respect to the
direct measurements) as we can immediately see also in Table
\ref{tbl:other_ckm_par}. Except for the fits of last year, all the
angles of the UT fits have nevertheless being in agreement with
that condition at $68\%$ C.L. But there is still the possibility
that there could be a sizable beyond the SM  contribution that
could show in future analyses, therefore changing the contribution
of the SM values to the fitted values of the unitary angles.
Note also that the central value of the direct experimental value
of $\alpha$ is lower than the central value of the CKM fitter and
the errors are comparable.

Another thing to consider, of course, is that the model based in
Eqs.~(\ref{eq:sym_y11_text}), (\ref{eq:mix_angl_th}) and
(\ref{eq:angles_st__forckm}) would need to include corrections or
modifications. Among them are the supersymmetric corrections to
the quark masses that should be carefully taken into account, or
deviations from the symmetric textures. These deviations are
indeed formally present since the symmetric structure of the mass
matrices is valid just at the GUT scale and may get sizable
modification by the RGE running to the  electroweak scale. Given
the increasing precision in the determination of the unitary
triangle fits, this running should be taken into account. The
running effects give a correction to the relation $s^u_{12}$ of
the form $s^u_{12}\approx 1/r \sqrt{m_u/m_c}$, where $r$ is a
parameter of $O(1)$ measuring the slight non-symmetry of the
elements $|Y^u_{12}|$ and $|Y^u_{21}|$. Another possible
modification is the one pointed out in \cite{Masina:2006ad}, namely
allowing the contribution of $Y^u_{11}$ to become non negligible.
This produces the same relation of  $s^u_{12}\approx 1/r
\sqrt{m_u/m_c}$ with $r$ depending on the non negligible
$Y^u_{11}$ element.    A separate question, independent of the fit
itself, is whether this fit is compatible with a particular
realization of a horizontal symmetry, like the one proposed in
\cite{Ross:2004qn}.

%

%

\begin{thebibliography}{99}

\bibitem{FY}
 M. Fukugita and T. Yanagida,
 Phys.\ Lett. {\bf B 174}, 45 (1986).



\bibitem{Ross:2002fb}
  G.~G.~Ross and L.~Velasco-Sevilla,
  Nucl.\ Phys.\ B {\bf 653}, 3 (2003)
  [arXiv:hep-ph/0208218].




\bibitem{Chen:2004xy}
  M.~C.~Chen and K.~T.~Mahanthappa,
  Phys.\ Rev.\ D {\bf 70}, 113013 (2004)
  [arXiv:hep-ph/0409096].
  M.~C.~Chen and K.~T.~Mahanthappa,
  Phys.\ Rev.\ D {\bf 65}, 053010 (2002)
  [arXiv:hep-ph/0106093].

\bibitem{Bando:2004hi}
  M.~Bando, S.~Kaneko, M.~Obara and M.~Tanimoto,
  arXiv:hep-ph/0405071.

\bibitem{Davidson:2002qv}
  S.~Davidson and A.~Ibarra,
  Phys.\ Lett.\ B {\bf 535}, 25 (2002)
  [arXiv:hep-ph/0202239].

\bibitem{Dermisek:2006dc}
  R.~Dermisek, M.~Harada and S.~Raby,
  Phys.\ Rev.\ D {\bf 74}, 035011 (2006)
  [arXiv:hep-ph/0606055].

\bibitem{Dermisek:2005ij}
  R.~Dermisek and S.~Raby,
  Phys.\ Lett.\ B {\bf 622}, 327 (2005)
  [arXiv:hep-ph/0507045].

\bibitem{Maekawa:2001uk}
  N.~Maekawa,
  Prog.\ Theor.\ Phys.\  {\bf 106}, 401 (2001)
  [arXiv:hep-ph/0104200].

\bibitem{Shafi:2005rd}
  Q.~Shafi and Z.~Tavartkiladze,
  Phys.\ Lett.\ B {\bf 633}, 595 (2006)
  [arXiv:hep-ph/0509237].

\bibitem{Berezhiani:2000cg}
  Z.~Berezhiani and A.~Rossi,
  Nucl.\ Phys.\ B {\bf 594}, 113 (2001)
  [arXiv:hep-ph/0003084].

\bibitem{Kitano:2000xk}
  R.~Kitano and Y.~Mimura,
  Phys.\ Rev.\ D {\bf 63}, 016008 (2001)
  [arXiv:hep-ph/0008269].

\bibitem{Chen:2003zv}
  M.~C.~Chen and K.~T.~Mahanthappa,
  Int.\ J.\ Mod.\ Phys.\ A {\bf 18}, 5819 (2003)
  [arXiv:hep-ph/0305088].



\bibitem{Campbell:1992hd}
  B.~A.~Campbell, S.~Davidson and K.~A.~Olive,
  Nucl.\ Phys.\ B {\bf 399}, 111 (1993)
  [arXiv:hep-ph/9302223].

\bibitem{Mohapatra:2005wg}
  R.~N.~Mohapatra {\it et al.},
  arXiv:hep-ph/0510213.

\bibitem{D'Ambrosio:2003wy}
 Y. Grossman, T. Kashti, Y. Nir and E. Roulet,
 Phys.\ Rev.\ Lett.\ {\bf 91}, 251801 (2003) [arXiv:hep-ph/0307081];
  G.~D'Ambrosio, G.~F.~Giudice and M.~Raidal,
  Phys.\ Lett.\ B {\bf 575}, 75 (2003)
  [arXiv:hep-ph/0308031].

\bibitem{Gonzalez-Garcia:2004jd}
  M.~C.~Gonzalez-Garcia,
  Phys.\ Scripta {\bf T121}, 72 (2005)
  [arXiv:hep-ph/0410030].


\bibitem{Mohapatra:1986bk}
  R. Mohapatra. ``Unification and Supersymmetry''. Springer Verlag 1986.

\bibitem{Fukuda:1998fd}
  Y.~Fukuda {\it et al.}  [Super-Kamiokande Collaboration],
  Phys.\ Rev.\ Lett.\  {\bf 81}, 1158 (1998)
  [Erratum-ibid.\  {\bf 81}, 4279 (1998)]
  [arXiv:hep-ex/9805021].




\bibitem{Albright:2002np}
  C.~H.~Albright,
  Int.\ J.\ Mod.\ Phys.\ A {\bf 18}, 3947 (2003)
  [arXiv:hep-ph/0212090].


\bibitem{Roberts:2001zy}
  R.~G.~Roberts, A.~Romanino, G.~G.~Ross and L.~Velasco-Sevilla,
  Nucl.\ Phys.\ B {\bf 615}, 358 (2001)
  [arXiv:hep-ph/0104088].


\bibitem{Kane:2005va}
  G.~L.~Kane, S.~F.~King, I.~N.~R.~Peddie and L.~Velasco-Sevilla,
  JHEP {\bf 0508}, 083 (2005)
  [arXiv:hep-ph/0504038].




\bibitem{Gatto:1968ss}
R. ~Gatto, G. ~ Sartori and M. ~Tonin, Phys.\ Lett. \ B {\bf 28},
128 (1968).


\bibitem{Ross:2004qn}
  G.~G.~Ross, L.~Velasco-Sevilla and O.~Vives,
  Nucl.\ Phys.\ B {\bf 692}, 50 (2004)
  [arXiv:hep-ph/0401064].




\bibitem{Masina:2006ad}
  I.~Masina and C.~A.~Savoy,
  Nucl.\ Phys.\  B {\bf 755} (2006) 1
  [arXiv:hep-ph/0603101].

\bibitem{Masina:2006pe}
  I.~Masina and C.~A.~Savoy,
  Phys.\ Lett.\  B {\bf 642}, 472 (2006)
  [arXiv:hep-ph/0606097].



\bibitem{Altarelli:2000fu}
  G.~Altarelli, F.~Feruglio and I.~Masina,
  JHEP {\bf 0011}, 040 (2000)
  [arXiv:hep-ph/0007254].



\bibitem{Anderson:1993fe}
  G.~Anderson, S.~Raby, S.~Dimopoulos, L.~J.~Hall and G.~D.~Starkman,
  Phys.\ Rev.\ D {\bf 49}, 3660 (1994)
  [arXiv:hep-ph/9308333].


\bibitem{Buch}
 For a review, see,
 W. Buchm\"uller, P. Di Bari and M. Pl\"umacher,
 arXiv:hep-ph/0401240.

\bibitem{scopel}
 E. J. Chun and S. Scopel, Phys.\ Lett.\ B {\bf 636},  278 (2006) [arXiv:hep-ph/0510170].

\bibitem{bbn1}
J.~R.~Ellis, A.~D.~Linde and D.~V.~Nanopoulos,
 Phys.\ Lett.\ B {\bf 118}, 59 (1982);
 M.~Y.~Khlopov and A.~D.~Linde, Phys.\ Lett.\ B {\bf 138}, 205 (1984);
 J.~R.~Ellis, J.~E.~Kim, and D.~V.~Nanopoulos,
 Phys.\ Lett.\ B {\bf 145}, 181 (1984);
 K.~Kohri, Phys.\ Rev.\ D {\bf 64} 043515 (2001) [arXiv:astro-ph/0103411];
 M. Kawasaki, K. Kohri and T. Moroi,  Phys.\ Lett.\ B {\bf 625}, 7 (2005)
 [arXiv:astro-ph/0402490].

\bibitem{bbn2}
 F.~D.~Steffen, JCAP {\bf 0609}, 001 (2006) [arXiv:hep-ph/0605306];
 J.~Pradler and F.~D.~Steffen, Phys.\ Rev.\ D {\bf 75} 023509 (2007)
[arXiv:hep-ph/0608344].


\bibitem{Yamaguchi:2002zy}
  M.~Yamaguchi and K.~Yoshioka,
  Phys.\ Lett.\ B {\bf 543},  189 (2002)
  [arXiv:hep-ph/0204293].



\bibitem{Giudice:1988yz}
  G.~F.~Giudice and A.~Masiero,
  Phys.\ Lett.\ B {\bf 206},  480 (1988).

\bibitem{Kim:1994eu}
  J.~E.~Kim and H.~P.~Nilles,
  Mod.\ Phys.\ Lett.\ A {\bf 9},  3575 (1994)
  [arXiv:hep-ph/9406296].


\bibitem{chun}
 E. J. Chun,
 Phys.\ Rev.\ {\bf D 69}, 117303 (2004) [arXiv:hep-ph/0404029].

\bibitem{Brignole}
 For a review, see,
 A. Brignole, L. E. Ibanez and C. Munoz,
 arXiv:hep-ph/9707209.


\bibitem{Pierce:1996zz}
  D.~M.~Pierce, J.~A.~Bagger, K.~T.~Matchev and R.~J.~Zhang,
  Nucl.\ Phys.\ B {\bf 491}, 3 (1997)
  [arXiv:hep-ph/9606211].

\bibitem{pdg07} W.-M. Yao et al.,  Journal of Physics G 33, 1 (2006).

\bibitem{ckmfitt_coll} http://www.slac.stanford.edu/xorg/ckmfitter/

\bibitem{ckmf07} http://www.slac.stanford.edu/xorg/ckmfitter/plots\_beauty06/ckmEval\_results\_beauty06.ps.gz

\bibitem{utFit_coll} http://utfit.roma1.infn.it/

\bibitem{Velasco-Sevilla:2006dy}
  L.~Velasco-Sevilla,
  arXiv:hep-ph/0603115.

\bibitem{Senoguz:2007hu}
  V.~N.~Senoguz,
  arXiv:0704.3048 [hep-ph].

\end{thebibliography}
\end{document}